\documentclass[conference]{IEEEtran}
\IEEEoverridecommandlockouts

\usepackage[numbers]{natbib}
\usepackage[nocompress]{cite}

\usepackage{microtype}
\usepackage{graphicx}
\usepackage{booktabs} %

\usepackage{amsmath,amsfonts,bm}

\def\eqref#1{equation~\ref{#1}}

\def\1{\bm{1}}

\DeclareMathAlphabet{\mathsfit}{\encodingdefault}{\sfdefault}{m}{sl}
\SetMathAlphabet{\mathsfit}{bold}{\encodingdefault}{\sfdefault}{bx}{n}

\newcommand{\E}{\mathbb{E}}

\newcommand{\R}{\mathbb{R}}

\DeclareMathOperator{\sign}{sign}

\PassOptionsToPackage{hyphens}{url}\usepackage[hidelinks]{hyperref}

\usepackage{xspace}
\usepackage{xfrac}
\usepackage{subcaption}
\usepackage{caption}
\usepackage[dvipsnames]{xcolor}
\usepackage{float}

\usepackage{amsmath}
\usepackage{amssymb}
\usepackage{mathtools}
\usepackage{amsthm}
\usepackage{bbm}
\usepackage{multirow}
\usepackage{enumitem}
\usepackage{tikz}
\usepackage{nicefrac}
\usetikzlibrary{shapes}
\usetikzlibrary{shapes.misc}
\tikzset{cross/.style={cross out, draw=black, minimum size=2*(#1-\pgflinewidth), inner sep=0pt, outer sep=0pt},
cross/.default={1pt}}
\usepackage{standalone}
\usepackage[most]{tcolorbox}
\usepackage{tabularx}

\newcommand{\opt}{\textsc{Opt}\xspace}
\newcommand{\signopt}{\textsc{Sign-Opt}\xspace}
\newcommand{\hsj}{\textsc{HopSkipJump}\xspace}
\newcommand{\rays}{\textsc{RayS}\xspace}
\newcommand{\ba}{\textsc{Boundary Attack}\xspace}

\newcommand{\stealthyopt}{\textsc{Stealthy Opt}\xspace}
\newcommand{\stealthysignopt}{\textsc{Stealthy Sign-Opt}\xspace}
\newcommand{\stealthyrays}{\textsc{Stealthy RayS}\xspace}
\newcommand{\stealthyhsj}{\textsc{Stealthy HSJA}\xspace}

\newcommand{\dist}{\texttt{dist}\xspace}
\renewcommand{\sign}{\texttt{sign}\xspace}
\newcommand{\getdist}{{\color{ForestGreen}\texttt{getDist}}\xspace}
\newcommand{\checkadv}{{\color{red}\texttt{checkAdv}}\xspace}

\newcommand{\findbound}{\texttt{projBoundary}\xspace}
\newcommand{\updatedir}{\texttt{updateDir}\xspace}
\newcommand{\stepsize}{\texttt{stepSize}\xspace}
\newcommand{\xmisclass}{\hat{x}}
\newcommand{\xboundary}{x_b}

\newcommand{\specialcell}[2][c]{%
  \begin{tabular}[#1]{@{}c@{}}#2\end{tabular}}

\makeatletter
\def\@IEEEsectpunct{.\ \,}
\def\paragraph{\@startsection{paragraph}{4}{\z@}{1.5ex plus 1.5ex minus 0.5ex}%
{0ex}{\normalfont\normalsize\bfseries}}
\makeatother

\newtheorem{theorem}{Theorem}

\newtcbtheorem{Summary}{}{enhanced,
  coltitle=black,
  top=0.14in,
  attach boxed title to top left=
  {xshift=0em,yshift=-\tcboxedtitleheight/2},
  boxed title style={size=small,colback=blue!10}
}{summary}

\newcommand{\flagged}{flagged\xspace}
\newcommand{\Flagged}{Flagged\xspace}
\newcommand{\nonflagged}{non-flagged\xspace}
\newcommand{\Nonflagged}{Non-flagged\xspace}

\newcommand{\plotheight}{4cm}

\usepackage{wrapfig}

\usepackage[capitalize,noabbrev]{cleveref}

\begin{document}

\title{Evading Black-box Classifiers\\ Without Breaking Eggs\\
\thanks{Edoardo Debenedetti is supported by armasuisse Science and Technology.}
}

\author{Edoardo Debenedetti$^1$ \qquad Nicholas Carlini$^2$ \qquad Florian Tramèr$^1$ \\
\emph{$^1$ETH Zurich \qquad $^2$Google DeepMind}}

\maketitle

\begin{abstract}
	Decision-based evasion attacks repeatedly query a black-box classifier to generate adversarial examples.
	Prior work measures the cost of such attacks by the total number of queries made to the classifier. We argue this metric is incomplete. Many security-critical machine learning systems aim to weed out ``bad'' data (e.g., malware, harmful content, etc). Queries to such systems carry a fundamentally \emph{asymmetric cost}: \emph{flagged queries}, i.e., queries detected as ``bad'' by the classifier come at a higher cost because they trigger additional security filters, e.g., usage throttling or account suspension. Yet, we find that existing decision-based attacks issue a large number of queries that would get flagged by a security-critical system, which likely renders them ineffective against such systems. We then design new attacks that reduce the number of flagged queries by $\mathbf{1.5}$--$\mathbf{7.3\times}$.
    While some of our attacks achieve this improvement at the cost of only a moderate increase in total (including non-flagged) queries, other attacks require significantly more total queries than prior attacks.
    We thus pose it as an open problem to build black-box attacks that are more effective under realistic cost metrics.
\end{abstract}

\begin{IEEEkeywords}
 security, threat models, black-box adversarial examples, decision-based attacks
\end{IEEEkeywords}

\section{Introduction}

Adversarial examples~\citep{szegedy2013intriguing, biggio2013evasion} are a security risk for machine learning (ML) models that interact with malicious actors.
For example, an attacker could use adversarial examples to post undesired content to the Web while bypassing ML filtering mechanisms~\citep{tramer2018adblock, prokos2023squint, rosenberg2021adversarial}.
In such security-critical uses of ML, the attacker often only has \emph{black-box} access to the ML model's decisions. %

\emph{Decision-based attacks}~\citep{brendel2017decision} generate adversarial examples in black-box settings by repeatedly querying the model and observing only the output decision on perturbed inputs. The original \ba of~\citet{brendel2017decision} required over 100,000 model queries to reliably find small adversarial perturbations. Subsequent work~\citep{cheng2018query, cheng2019sign, chen2020hopskipjumpattack, chen2020rays, li2020qeba} has optimized for this metric of ``total number of model queries'', and reduced it by $1$--$3$ orders of magnitude.

\textbf{We argue this metric fails to reflect the true cost of querying a security-critical ML system}. Such systems typically aim to detect ``bad'' data, such as malware, harmful content or malicious traffic. Queries with benign data (e.g., a selfie uploaded to social media) carry little cost; in contrast, bad data that is detected and flagged by the system (e.g., offensive content) triggers additional security measures that carry a high cost for the attacker---up to account termination~\citep{youtube,facebook}.
Thus, we argue that black-box attacks should strive to be \emph{stealthy}, by minimizing the number of ``bad'' queries that are flagged by the ML system---rather than the total number of queries (both bad and benign).

\begin{figure}[t]
	\centering
	\includegraphics[width=\columnwidth]{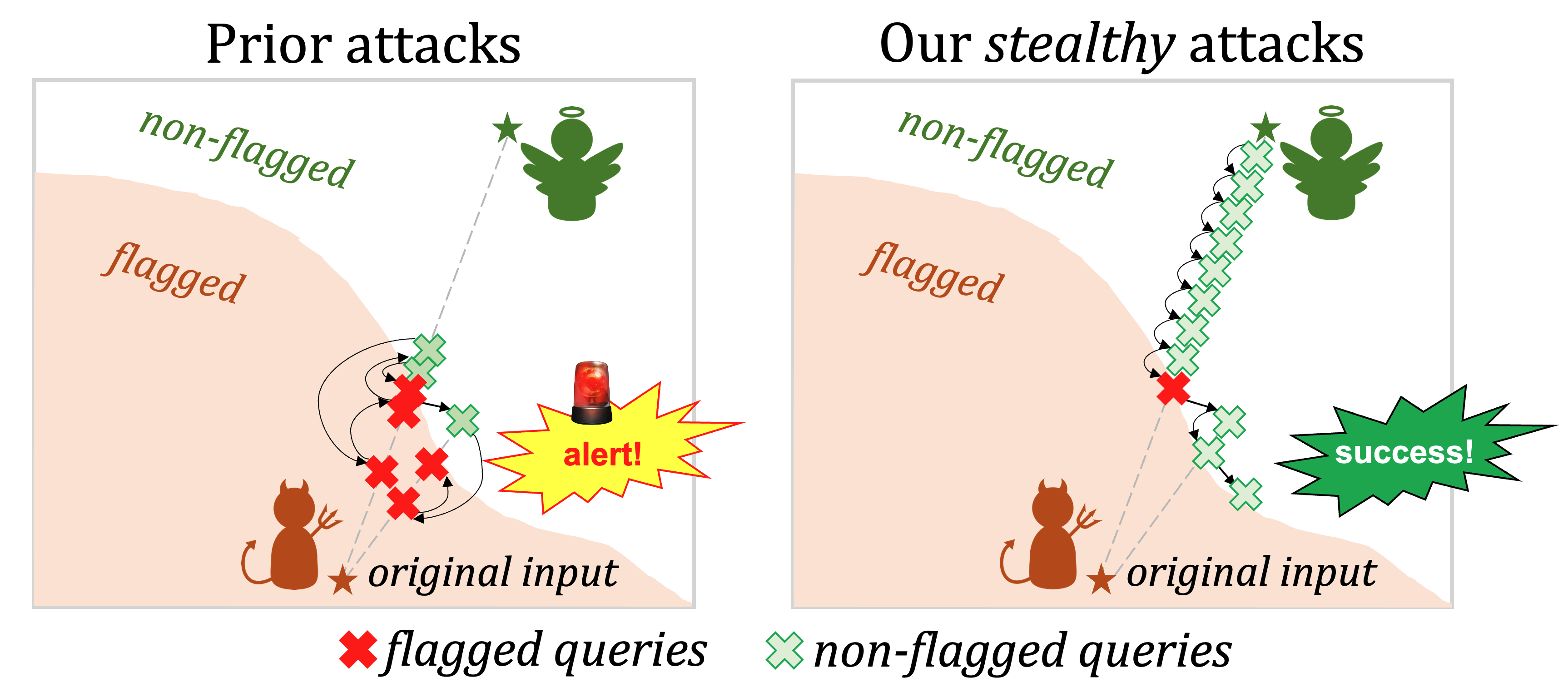}
	\caption{Existing decision-based attacks (left) make many ``flagged'' queries, that get classified into the class that the attacker aims to evade (e.g., NSFW content). In security-critical applications, such flagged queries would trigger additional security mechanisms, thus raising the cost of the attack. Our \emph{stealthy} attacks (right) trade-off \flagged{} queries for \nonflagged{} ones, to find adversarial examples without raising security alerts.}
	\label{fig:figure1}
    \vspace{-0.5cm}
\end{figure}

We find that existing attacks are not stealthy: over 50\% of the queries they make are \flagged.
We then show how to drastically reduce the number of \flagged{} queries for a class of attacks that measure distances to the model's boundary along random directions (e.g., \opt~\citep{cheng2018query}, \signopt~\citep{cheng2019sign} and \rays~\citep{chen2020rays}).
Inspired by the famous ``egg-dropping problem''\footnote{The egg-dropping problem is a mathematical exercise that asks to find the maximal storey in a building from which an egg can be dropped without breaking while incurring at most $k \geq 1$ broken eggs: \url{https://brilliant.org/wiki/egg-dropping}.},
we design variants of these attacks (for both the $\ell_\infty$ and $\ell_2$ norms) that trade-off \flagged{} queries for \nonflagged ones.

We evaluate our attacks on three classification tasks: ImageNet, a binary dog vs.~not-dog classification task on ImageNet, and NSFW content detection~\citep{schuhmann2022laion}. %
Our stealthy attacks reduce the number of \flagged{} queries of the original attacks by $1.5$--$7.3\times$. Notably, our stealthy variant of the $\ell_\infty$ \rays attack reduces \flagged{} queries by $2.1$--$2.5\times$ over \rays and $6$--$17\times$ over \hsj, while making $2.1$--$3.4\times$ more benign, \nonflagged{} queries. We then use our stealthy \rays attack to evade a commercial black-box NSFW image detector, with $2.2\times$ fewer \flagged{} queries than the original \rays attack.

For $\ell_2$-attacks, our stealthy attacks similarly reduce the number of \flagged{} queries compared to prior attacks.
Notably, our stealthy variant of the \opt attack outperforms \signopt and \hsj in terms of \emph{\flagged} queries, despite the two latter attacks issuing fewer queries \emph{in total}.
However, our stealthy attacks also incur a much higher cost 
in \nonflagged{} queries (and thus in the number of total queries issued by the attack).
Concretely, our most stealthy $\ell_2$ attack (based on \hsj) reduces \flagged{} queries by a factor $1.5$--$1.8\times$, but incurs a large increase in \nonflagged queries ($350$--$1{,}400\times$).
Our stealthy $\ell_2$ attacks are thus likely only cost-effective in scenarios where \flagged queries carry a significantly higher cost ($\approx 1{,}000\times$) than \nonflagged ones (this cost ratio may be realistic when \flagged queries are likely to trigger costly measures such as account termination).

Overall, our results suggest that many decision-based attacks are far from stealthy and that
stealthier attacks are viable if the cost of \flagged{} queries far outweighs that of \nonflagged{} queries (especially for $\ell_2$ attacks). %
We thus recommend that future decision-based attacks account for asymmetric query costs, to better reflect the true cost of deploying such attacks against real security-critical systems.

\section{Decision-based Attacks}
\label{sec:background}

Given a classifier $f: [0,1]^d \to \mathcal{Y}$ and input $(x,y)$, an (untargeted) adversarial example $\hat{x}$ is an input close to $x$ that is misclassified, i.e., $f(\hat{x}) \neq y$ and $\|\hat{x}- x\|_p \leq \epsilon$ for some $\ell_p$ norm and threshold $\epsilon$.

A decision-based attack gets oracle access to the model $f$. The attacker can query the model on arbitrary inputs $x \in [0,1]^d$ to obtain the class label $y \gets f(x)$.
Existing decision-based attacks
aim to minimize the \textbf{total number of queries} made to the model $f$ before the attack succeeds.

\paragraph{Applications}
Decision-based attacks~\citep{brendel2017decision} were designed for black-box ML systems that only return model decisions (e.g., an ML model that filters social media content).
Such attacks are also applicable when an attacker has physical access to a model guarded by hardware protections, e.g., a phone's authentication mechanism, or a self-driving system.
Decision-based attacks are also commonly used to evaluate the robustness of \emph{white-box} models, when computing gradients is hard~\citep{brendel2017decision, tramer2020adaptive}.

In this paper, we are interested in the first two scenarios, where decision-based attacks are used against black-box ML security systems. In particular, we assume that these security systems monitor and log user queries, and can throttle or disable an attacker's access to the system.

\section{Asymmetric Query Costs}
\label{sec:bad_queries}

Existing decision-based attacks optimize for the \emph{total} number of model queries.
This is reasonable if the attacker's primary cost is incurred by queries to the model, and this cost is \emph{uniform} across queries (e.g. if the attacker has to pay a fixed service fee for each query).

However, we argue that query costs are rarely uniform in practical security-critical systems. This is because, in such systems, the goal of an ML model is usually to detect and flag ``bad'' data (e.g., malware, harmful content, malicious traffic, etc).
The costs incurred by querying such a model are highly asymmetric. Querying the model with ``good'' data that does not get flagged is expected, comes with no additional overhead, and is thus cheap.
Whereas querying the model with ``bad'' data is unexpected, triggers additional security measures and filters, and thus places a much higher cost on the attacker.

We first formalize the notion of ``flagged'' and ``non-flagged'' queries. 
We assume that the ML system uses a classifier $f: [0,1]^d \to \mathcal{Y}$ to filter out bad content, and that some subset of the output classes $\mathcal{Y}_{\textrm{bad}} \subset \mathcal{Y}$ correspond to bad data (e.g., social media content that is NSFW, offensive, etc.) We then define \textit{\flagged} and \textit{\nonflagged} queries (or inputs) as follows:

\begin{Summary}[title=\textbf{\Flagged queries}]{}{}
	A query $x$ is \textit{\flagged} if the target model $f$ labels it as bad so that it is filtered out, i.e., $f(x) \in \mathcal{Y}_{\textrm{bad}}$.
\end{Summary}

Conversely:

\begin{Summary}[title=\textbf{\Nonflagged queries}]{}{}
    A query $x$ is \textit{\nonflagged} if the target model classifies it as benign, i.e., $f(x) \notin \mathcal{Y}_{\textrm{bad}}$. That is, if it is not \textit{\flagged}.
\end{Summary}

It is important to note that a query $x$ being \flagged or \nonflagged is solely a property of the classifier $f$ and \emph{not} of the input's ground truth label. For example, a benign image (e.g., a cute puppy) can be flagged; this is a \emph{false-positive} of the classifier. Conversely, a \textbf{successful} adversarial example for an NSFW image will---by definition---be \textbf{\nonflagged{}} even though it is objectively bad; this is thus a \emph{false-negative} of the classifier.

In an evasion attack, the attacker is given an input $(x,y)$ that is (objectively) bad, i.e., $y \in \mathcal{Y}_{\textrm{bad}}$, 
and their goal is to find an adversarial example $\hat{x}$ that is not flagged by the system, i.e., $f(\hat{x}) \notin \mathcal{Y}_{\textrm{bad}}$.
All queries made by the attacker to the model $f$ carry a base cost $c_0$, due to data processing, network bandwidth, disk storage, or throttling if the attacker makes too many queries.
This base cost is typically very low: e.g., Facebook users can upload $1{,}000$ images at once in an album~\citep{facebook_albums}.
However, for queries $x$ that are flagged as inappropriate (i.e., $f(x) \in \mathcal{Y}_{\textrm{bad}}$), the cost $c_{\textrm{\flagged}}$ incurred by the attacker is much larger. Their account could be suspended or banned, their IP blacklisted, etc. While these restrictions can be circumvented (e.g., by buying multiple accounts~\citep{facebook_accounts}), this places a significantly higher cost on queries flagged as \flagged, i.e., $c_{\textrm{\flagged}} \gg c_0$.

We thus argue that decision-based attacks should strive to minimize the following cost:%
\begin{equation}
	\textrm{\emph{minimize}}\quad\texttt{cost} \coloneqq Q_{\textrm{total}} \cdot c_0 + Q_{\textrm{\flagged}} \cdot c_{\textrm{\flagged}} \;,
\end{equation}%
where $Q_{\textrm{\flagged}}$ is the number of \flagged{} model queries ($f(x) \in \mathcal{Y}_{\textrm{\flagged}}$), $Q_{\textrm{total}}$ is the total number of queries---including \flagged{} ones---and $c_{\textrm{\flagged}}\gg c_0$.
We call attacks that minimize this asymmetric cost \emph{stealthy}.

\paragraph{Example: evading NSFW content detection}
To make the above discussion more concrete, consider the example 
of an attacker who tries to upload inappropriate content to a social media website. Every uploaded image passes through an ML model that flags inappropriate content. Here, a \textit{\flagged} query is \textbf{any} query to the system that the NSFW detector classifies as NSFW, and which will thus be filtered out.

Uploading content that is \nonflagged carries little cost: the social media platform will simply post the picture, and only apply rate limits once the user uploads a very large number of \nonflagged{} content. But if a query \emph{is} \flagged as inappropriate, the system is likely to block the contents and take further costly actions (e.g., account throttling, suspension, or termination). 
We thus have that $c_{\textrm{\flagged}}\gg c_0$ (we further discuss how to estimate these costs below and in \Cref{sec:costs}).

\paragraph{Existing attacks are not designed to be stealthy}
No existing black-box attack considers such asymmetric query costs. As a result, these attacks issue a large number of \flagged{} queries.
We illustrate this with an untargeted attack on ImageNet.\footnote{ImageNet is not a security-critical task, and thus most content is not ``bad''. We use ImageNet here because prior attacks were designed to work well on it. To mimic a security-critical evasion attempt, we set the class to be evaded as ``\flagged'' and all other classes as ``\nonflagged''. That is, for an input $(x, y)$ we set $\mathcal{Y}_{\textrm{\flagged}} = \{y\}$ and the attacker's goal is to find an adversarial example $\hat{x}$ such that $f(\hat{x}) \neq y$ while avoiding making queries labeled as $y$.}
In \Cref{tab:bad_queries}, we show the number of total queries $Q_{\textrm{total}}$ and \flagged{} queries $Q_{\textrm{\flagged}}$ made by various $\ell_2$ and $\ell_\infty$ decision-based attacks on a ResNet-50 classifier.
In all cases, half or more of the attacker's queries are ``\flagged'' (i.e., they get the class label that was to be evaded).
Despite differences in the fraction of \flagged{} queries for each attack, attacks that make fewer total queries  also make fewer \flagged{} queries. But this begs the question of whether we could design attacks that issue far fewer \flagged{} queries in total. The remainder of this paper answers this question.

\begin{table}[t]
        \normalsize
	\caption{Median number of queries for each attack to reach a median $\ell_2$ distance of 10 and median $\ell_\infty$ distance of $\sfrac{8}{255}$ on untargeted ImageNet. We report the total number of attack queries $Q_{\textrm{total}}$, and of ``\flagged'' queries $Q_{\textrm{\flagged}}$ (queries that get classified as the  class that the attacker wants to evade).}
	\vspace{0.75em}
	\centering
	\setlength{\tabcolsep}{4pt}
	\begin{tabular}{@{}llrr@{}}
		&&
		\textbf{Total Queries}  &
		\textbf{\Flagged{} Queries} 
		\\
            \textbf{Norm}                               &
		\textbf{Attack}                             &
            ($Q_{\textrm{total}}$)
            &
            ($Q_{\textrm{\flagged}}$)
            \\
		\toprule
		\multirow{4}{*}{$\ell_2$}                   & \opt              & $9{,}731$ & $4{,}975$ (51\%) \\
		                                            & \textsc{Boundary} & $4{,}555$ & $3{,}843$ (84\%) \\
		                                            & \signopt          & $2{,}873$ & $1{,}528$ (53\%) \\
		                                            & \hsj              & $1{,}752$ & $953$ (54\%)     \\ \midrule
		\multirow{2}{*}{$\ell_\infty$}              & \hsj              & $3{,}591$ & $1{,}789$ (50\%) \\
		                                            & \rays             & $328$     & $244$ (74\%)     \\ \bottomrule
	\end{tabular}
	\label{tab:bad_queries}
\end{table}

\paragraph{Selecting the values of $c_0$ and $c_{\mathrm{\flagged}}$}
The true cost of a query (whether \nonflagged{} or \flagged) may be hard to estimate, and can vary between applications.
As a result, we recommend that black-box attack evaluations report both the value of $Q_{\textrm{total}}$ and $Q_{\textrm{\flagged}}$ so that the attack cost can be calculated for any domain-specific values of $c_0$ and $c_{\textrm{\flagged}}$.

In this paper, we often make the simplifying assumption that $c_0=0, c_{\textrm{\flagged}}=1$, a special case that approximates the attack cost when $c_{\textrm{\flagged}} \gg c_0$.
In this special case, the attacker solely aims to minimize \flagged{} queries, possibly at the expense of a large increase in total queries.
We will, however, also consider what the trade-offs look like in real-world applications in \Cref{sec:costs}.

\section{Designing Stealthy Decision-based Attacks}
\label{sec:attacks}

We explore the design space of stealthy decision-based attacks, which minimize the total number of \flagged{} queries made to the model.

One possibility is simply to design a \emph{better} decision-based attack, that makes fewer \emph{total} queries. As we see from \Cref{tab:bad_queries}, this is how prior work has implicitly minimized asymmetric attack costs so far.
We take a different approach, and design attacks that explicitly \emph{trade-off} \flagged{} queries for \nonflagged{} ones.

In \Cref{ssec:decision_attacks}, we first review how existing decision-based attacks work, and distill some common sub-routines that help us understand how these attacks spend their queries (either \flagged{} or \nonflagged{} ones).
In \Cref{ssec:eggdrop}, we then show how to design stealthy variants of these sub-routines, and in \Cref{ssec:our_attacks} we describe the resulting stealthy attacks that we introduce.
In \Cref{subsec:theory} we formally prove that our stealthy attack techniques are more efficient (in terms of \flagged{} queries) than prior attacks.

\subsection{How do Decision-based Attacks Work?}
\label{ssec:decision_attacks}

\begin{table*}[ht!]
	\centering
        \normalsize
	\caption{Queries issued by different decision-based attacks in a single attack iteration.
		We distinguish between \checkadv queries that check whether some arbitrary direction yields a misclassification and \getdist queries that issue multiple calls to the model to measure the distance to the model boundary along some direction.
		The hyper-parameter $n$ is the number of times a routine is called for estimating the geometry of the model's decision boundary. The variable $m$ is the average number of step-size searches conducted in one iteration of \opt and \signopt.}
	\vspace{0.5em}
	\setlength{\tabcolsep}{4pt}
	\begin{tabular}{@{}l l l l c|c l c l @{}}
		                    & \multicolumn{3}{c}{\textbf{Attack Phase}}                                                                 \\
		\cmidrule(l{0pt}r{0pt}){2-4}
		\textbf{Attack}
		                    & \textbf{\findbound}
		                    & \textbf{\updatedir}
		                    & \textbf{\stepsize}
		                    & \multicolumn{1}{c}{}                      &
		                    & \multicolumn{3}{c}{\textbf{Total}}                                                                        \\
		\toprule
		\textsc{Boundary}   & $\checkadv \cdot n$                       &
		--                  & $\checkadv$                               &   &  & $\checkadv \cdot (n+1)$ &   &                          \\
		\opt                & $\getdist$                                &
		$\getdist \cdot n$  & $\getdist \cdot m$                        &   &  &                         &   & $\getdist \cdot (n+m+1)$ \\
		\signopt            & $\getdist$                                &
		$\checkadv \cdot n$ & $\getdist \cdot m$                        &   &  & $\checkadv \cdot n$     & + & $\getdist \cdot (m+1)$   \\
		\hsj                & $\getdist$                                &
		$\checkadv \cdot n$ & $\getdist$                                &   &  & $\checkadv \cdot n$     & + & $\getdist \cdot 2$       \\
		\rays               & $\getdist$                                &
		--                  & $\checkadv$                               &   &  & $\checkadv$             & + & $\getdist$               \\
		\bottomrule
	\end{tabular}
	\label{tab:generic_attack_phases}
\end{table*}

Most decision-based attacks follow the same blueprint~\citep{fu2022autoda}. For an input $(x, y)$, the attacks first pick an \emph{adversarial direction} $\theta \in [0, 1]^d$ and find the $\ell_p$ distance to the model's decision boundary from $x$ along the direction $\theta$.
They then iteratively perturb $\theta$ to minimize the boundary distance along the new direction.
Each iteration of the attacks can be divided into three phases:

\begin{itemize}
	\item $\findbound$: given the original input $(x, y)$ and a search direction $\theta$, this phase finds a point $x_b$ that lies on the model's decision boundary along the line $x + \alpha \cdot \sfrac{\theta}{\|\theta\|}$., and returns the $\ell_p$ distance between $x$ and $x_b$, i.e., $\dist \gets \|x - \xboundary\|_p$.

	\item $\updatedir$: This phase searches for an update direction $\delta$ to be applied to the search direction $\theta$.

	\item $\stepsize$: This phase selects a step-size $\alpha$ for an update to the search direction $\theta$.
\end{itemize}

For our purposes, it will be helpful to further decompose each of these three phases into two fundamental subroutines:

\begin{itemize}
	\item $\getdist(x, \theta, p) \to \R^+$: this routine computes the distance (in $\ell_p$ norm) from $x$ to the decision boundary along the direction $\theta$. Most attacks do this by performing a binary search between $x$ and a misclassified point $\hat{x}$ in the direction $\theta$, up to some numerical tolerance $\eta$.
	\item $\checkadv(x, \theta', \dist, y) \to \{-1, 1\}$: this routine uses a single query to check if the point at distance $\dist$ in direction $\theta'$ is misclassified, i.e., it returns 1 if $f(x + \dist \cdot \sfrac{\theta'}{\|\theta'\|_p}) \neq y$.
\end{itemize}

Different attacks combine these two subroutines in different ways, as described below.
As we will see, how an attack balances these two routines largely impacts how stealthy the attack can be made. Briefly, calls to $\checkadv$ cannot be turned stealthy because each bad query provides just two bits of information to the attacker on average. In contrast, a call to $\getdist$ can be implemented so that a single bad query yields $\log \sfrac{1}{\eta}$ bits of information.

\paragraph{An overview of existing attacks}
We briefly review how different attacks make use of $\checkadv$ and $\getdist$ routines in the $\findbound$, $\updatedir$, and $\stepsize$ phases.\\

\noindent\ba:
The original decision-based attack of~\citet{brendel2017decision} is a greedy attack.
In contrast to other attacks, it only performs a heuristic, approximate projection to the model's boundary in each step.

\begin{itemize}
\item$\findbound$:
Given a misclassified point $x_b$ along the direction $\theta$ (originally a natural sample from a different class than $x$), the attack samples random points around $x_b$ and checks on which side of the boundary they fall. From this, the attack estimates a step size to project $x_b$ onto the boundary and then computes the distance $\dist$ between $x_b$ and $x$.
This requires $n$ calls to $\checkadv$.

\item$\updatedir$: The attack is greedy and simply picks a small update direction $\delta$ at random.

\item$\stepsize$: The attack checks whether the distance to the boundary along the new direction $\theta + \delta$ is smaller than the current distance, $\dist$. If not, the update is discarded. Note that this test can be performed with a single query to the model, with a call to $\checkadv$.\\
\end{itemize}

\begin{table*}[t]
        \normalsize
	\caption{Where do decision-based attacks spend their queries? We run untargeted attacks against a ResNet-50 on ImageNet (see \Cref{ssec:eval_setup} for details). For each attack, we report the fraction of queries used in $\checkadv$ or $\getdist$ routines, and the fraction of \flagged{} queries in each routine.}
	\vspace{0.25em}
	\centering
	\begin{tabular}{@{}l l r r r r@{}}
		                                       &
		                                       &
		\multicolumn{2}{c}{\textbf{\checkadv}} &
		\multicolumn{2}{c}{\textbf{\ \ \getdist}}                                                                                                           \\[-0.2em]
		\cmidrule(l{5pt}r{5pt}){3-4}\cmidrule(l{5pt}r{0pt}){5-6}
		\textbf{Norm}
		                                       &
		\textbf{Attack}
		                                       &
		\textbf{all}                           & \textbf{fraction \flagged} & \textbf{all}        & \textbf{fraction \flagged}                                    \\
		\toprule
		\multirow{4}{15pt}{$\ell_2$}
		                                       & \textsc{Boundary}       & {\color{red}100\%}  & 84\%                    & {\color{black}0\%}        & --   \\
		                                       & \opt                    & {\color{black}2\%}  & 50\%                    & {\color{ForestGreen}98\%} & 52\% \\
		                                       & \signopt                & {\color{red}77\%}   & 52\%                    & {\color{black}23\%}       & 57\% \\
		                                       & \hsj                    & {\color{red}93\%}   & 55\%                    & {\color{black}7\%}        & 43\% \\
		\midrule
		\multirow{2}{15pt}{$\ell_\infty$}
		                                       & \hsj                    & {\color{red}92\%}   & 50\%                    & {\color{black}8\%}        & 50\% \\
		                                       & \rays                   & {\color{black}36\%} & 67\%                    & {\color{ForestGreen}64\%} & 78\% \\
		\bottomrule
	\end{tabular}
	\label{tab:attack_phases}
 \vspace{-0.5cm}
\end{table*}

\noindent\rays:
This is a greedy attack similar to \ba, tailored to the $\ell_\infty$ norm. Its search direction $\theta \in \{-1, +1\}^d$ is always a signed vector.

\begin{itemize}
    \item$\findbound$: \rays find the current distance to the decision boundary using a binary search, by calling $\getdist$.

\item$\updatedir$: The attack picks a new search direction by flipping the signs of a all pixels in a rectangular region of $\theta$.

\item$\stepsize$: The attack greedily checks whether the new direction improves the current distance to the boundary, by issuing a call to $\checkadv$. If the distance is not reduced, the update is discarded.\\
\end{itemize}

\noindent\opt:
This attack first proposed a gradient-estimation approach to decision-based attacks.

\begin{itemize}
\item$\findbound$: The attack starts by measuring the distance to the boundary, with a call to $\getdist$. Specifically, it performs a binary search between $x$ and some point $\xmisclass$ of a different class along the direction $\theta$.

\item$\updatedir$: The attack estimates the gradient of the distance to the boundary along the search direction $\theta$.
To this end, it samples random directions $r_1, \dots, r_n$
and computes the distance to the boundary along $\theta + r_i$, denoted as $d_i \in \R^+$, for each. The estimated gradient is then:
\begin{equation}
	\delta \gets \frac{1}{n} \sum_{i=1}^{n} (\dist - d_i)\cdot r_i\;.
\end{equation}
The attack uses $n$ calls to $\getdist$ to compute the boundary distance along each random direction.

\item$\stepsize$: \opt computes the step-size $\alpha$ with a \emph{geometric search}: starting from a small step size, double it as long as this decreases the distance to the decision boundary along the new direction $\theta + \alpha\cdot \delta$. Thus, each step of the geometric search involves a call to $\getdist$.\\
\end{itemize}

\noindent\signopt and \hsj:
These attacks are very similar and improve over $\opt$ by using a more query-efficient gradient-estimation procedure.

\begin{itemize}
    \item$\findbound$: In \hsj, this step is viewed as a boundary ``projection'' step which returns the point $x_b$ on the boundary, while \signopt computes the  \emph{distance} from $x$ to the boundary along $\theta$. But the two views, and their implementations, are equivalent. Both attacks use a binary search to find a point $x_b$ on the boundary, as in \opt, with a call to $\getdist$.

\item$\updatedir$: Both attacks also sample $n$ random search directions $r_1, \dots, r_n$. But instead of computing the distance to the boundary along each updated direction as in \opt, \signopt, and \hsj simply check
whether each update decreases the current distance $\dist$ to the decision boundary or not. The update direction is computed as
\begin{equation}
	\delta \gets \frac{1}{n} \sum_{i=1}^{n} z_i \cdot r_i\;,
\end{equation}
where $z_i \in \{-1, +1\}$ is one if and only if the point at distance $\dist$ along the direction $\theta+r_i$ is misclassified. \hsj differs slightly in that the random directions $r_i$ are applied to the current point on the boundary $\xboundary$, and we check whether $\xboundary+r_i$ is misclassified or not.
Compared to \opt, these attacks thus only issue $n$ calls to $\checkadv$ (instead of $n$ calls to $\getdist$), but the gradient estimate they compute has a higher variance.

\item$\stepsize$: \signopt uses the exact same geometric step-size search as \opt. \hsj is slightly different from the generic algorithm described above, in that it applies the update $\delta$ to the current point on the boundary $\xboundary$. The attack starts from a large step size and halves it until $\xboundary + \alpha\cdot \delta$ is misclassified. This amounts to finding the distance to the boundary from $\xboundary$ along the direction $\delta$, albeit with a geometric backtracking search instead of a binary search.\\
\end{itemize}

\Cref{tab:generic_attack_phases} summarizes the calls made to \checkadv and \getdist by each attack. In \Cref{tab:attack_phases}, we show how many \flagged{} queries and total queries are used for both routines in an untargeted attack for a standard ResNet-50 on ImageNet (where we view the class to be evaded as ``\flagged'').

\subsection{Maximizing Information per \Flagged{} Query}
\label{ssec:eggdrop}
To design stealthy decision-based attacks, we first introduce the \emph{entropy-per-\flagged-query} metric. This is the information (measured in bits) that the attacker learns for every \flagged{} query made to the model.

Consider an attack that calls $\checkadv(x, \theta+r_i, \dist, y)$ for many random $r_i$.
\ba, \hsj and \signopt do this to estimate the shape of the decision boundary. For a locally linear boundary, we expect 50\% of such queries to be \flagged. The attacker thus learns two bits of information per \flagged{} query.
To increase the entropy-per-\flagged-query, we would need to sample the $r_i$ so that fewer $\checkadv$ queries are \flagged. But this requires a prior on the boundary's geometry, which is what these queries aim to learn. It thus seems hard to make this procedure stealthier.

For calls to \getdist, a standard binary search requires $\log \sfrac{1}{\eta}$ queries (half of which are \flagged) to estimate the boundary distance up to tolerance $\eta$. A call to \getdist thus gives $\log \sfrac{1}{\eta}$ bits of information.
So the attacker also learns an average of two bits per \flagged{} query. However, here there is a simple way to trade-off \flagged{} queries for \nonflagged{} ones, which lets the attacker learn the same $\log \sfrac{1}{\eta}$ bits of information with as little as one \flagged{} query. All that is required is a tall building and some eggs!

\paragraph{Measuring distances with one \flagged{} query}

In the famous ``egg-dropping problem'', there is a building of $N$ floors, and you need to find the highest floor $n \in [1, N]$ from which an egg can be dropped without breaking. The egg breaks if and only if dropped from above some unknown floor $n$.
In the simplest version of the problem, you have a single egg and must compute the value of $n$. The solution is to drop the egg from each floor consecutively starting from the first until it breaks.

We note that finding the decision boundary between $x$ and $\hat{x}$, while minimizing \flagged{} queries, is exactly the egg-dropping problem! Assuming $\|x-\hat{x}\|_p = 1$, a search tolerance of $\eta$ yields a ``building'' of $N=\sfrac{1}{\eta}$ ``floors'' of length $\eta$ from $\hat{x}$ to $x$. The first $n$ floors (up to the boundary) are \nonflagged{} queries, i.e., no broken egg. All floors above $n$ are \flagged{} queries on the wrong boundary side, i.e., a broken egg.

While a binary search minimizes the total number of queries for finding the boundary, a \emph{line-search}---which moves from $\hat{x}$ to $x$ until the boundary is hit---is optimal for minimizing \flagged{} queries.

Many attacks use a small search tolerance $\eta$ (on the order of $10^{-3}$), so a full line search incurs a large cost of \emph{\nonflagged} queries ($\sfrac{1}{\eta}$).
This tradeoff is warranted if \flagged{} queries are substantially more expensive than \nonflagged{} ones, i.e., $c_{\textrm{\flagged}} \gg \sfrac{c_0}{\eta}$.
We thus consider finer-grained methods to trade off \flagged{} and \nonflagged{} queries using the general version of the egg-dropping problem.

\begin{figure}
	\begin{center}
		\begin{tikzpicture}
\draw[gray, dotted] (0,1.2) -- (4,1.2);
\draw[gray, dotted] (0,0.6) -- (4,0.6);
\draw[gray, dotted] (0,0) -- (4,0);

\draw[gray, ultra thick]  (3.05, -0.15) to[out=90,in=-100] (3.05, 1.35);
\draw (-0.25, 1.2) node[anchor=east, align=right]{\footnotesize \emph{full line-search}};
\foreach \x in {0, 0.4, 0.8, 1.2, 1.6, 2, 2.4} {
    \draw[->, very thick] (\x,1.2) -- (\x+0.4,1.2) node{};
}
\draw[very thick] (2.8, 1.2) -- (3.2,1.2) node{};
\draw[thick] (2.8, 1.2) -- (3.2,1.2) node[cross,red,draw, scale=15]{};
\draw (0, 1.2) node[fill,star,star points=5,  star point height=.3cm, ForestGreen, scale=0.3, draw]{};
\draw (4.1, 1.2) node[fill,star,star points=5,  star point height=.3cm, brown, scale=0.3, draw]{};
\draw (-0.25, 0.6) node[anchor=east, align=right]{\footnotesize \emph{k-stage line-search}};
\foreach \x in {0, 0.9, 1.8} {
    \draw[->, very thick] (\x,0.6) -- (\x+0.9,0.6) node{};
}
\draw[very thick] (2.7, 0.6) -- (3.6, 0.6) node{};
\draw[thick] (2.7, 0.6) -- (3.6, 0.6) node[cross,red,draw, scale=15]{};
\draw (0, 0.6) node[fill,star,star points=5,  star point height=.3cm, ForestGreen, scale=0.3, draw]{};
\draw (4.1, 0.6) node[fill,star,star points=5,  star point height=.3cm, brown, scale=0.3, draw]{};

\draw[->, thin] (2.7, 0.6) -- (2.85, 0.6) node{};
\draw[->, thin] (2.85, 0.6) -- (3, 0.6) node{};
\draw (3, 0.6) -- (3.15, 0.6) node[cross,draw, red, scale=3]{};

\draw[thick, gray, dotted, ->]  (3.5, 0.7) to[out=135,in=45] (2.7, 0.7);

\draw (-0.25, 0) node[anchor=east, align=right]{\footnotesize \emph{early stopping}};
\foreach \x in {0, 0.4, 0.8, 1.2, 1.6, 2, 2.4} {
    \draw[->, very thick] (\x,0) -- (\x+0.4,0) node{};
}
\draw (0, 0) node[fill,star,star points=5,  star point height=.3cm, ForestGreen, scale=0.3, draw]{};
\draw (4.1, 0) node[fill,star,star points=5,  star point height=.3cm, brown, scale=0.3, draw]{};

\draw[<->, blue, dashed] (2.5, -0.2) -- node [below] {\footnotesize $\gamma\cdot\texttt{dist}$} (4.1, -0.2);

\end{tikzpicture}
	\end{center}
	\caption{Line-search strategies to find the boundary (in gray) between a benign input (green) and the original \flagged{} input (brown). Red crosses are \flagged{} queries.}
	\label{fig:linesearch}
\end{figure}
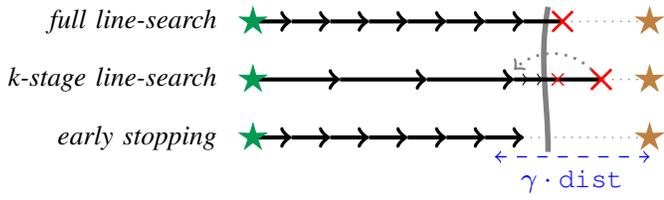
\paragraph{Trading \nonflagged{} and \flagged{} queries}
In the general version of the egg-dropping problem, you are given $k\geq 1$ eggs to find the safe height $n$ with a minimal number of egg drops.
This problem has a standard dynamic programming solution. 
Asymptotically, you need $\Theta(N^{\sfrac{1}{k}})$ egg drops given $k$ eggs, as we now show for $k=2$ eggs: first, divide the $N$ floors into $\sqrt{N}$ groups of $\sqrt{N}$ floors and do a
\emph{coarse-grained} line-search by dropping from floors $1, 1+\sqrt{N}, 1+2\sqrt{N}, \dots$ until the first egg breaks. You now know the solution is in the previous group
of $\sqrt{N}$ floors, so you do a \emph{fine-grained} line-search in this group one floor at a time. This requires at most $2\sqrt{N}$ egg drops.

For our boundary finding problem, we can thus divide the interval between $x$ and $\hat{x}$ into $\sfrac{1}{\eta}$ intervals, and do two line
searches with step-sizes respectively $\sqrt{\eta}$ and $\eta$. This will incur two \flagged{} queries, and $2\sqrt{\sfrac{1}{\eta}}$ total queries, compared to one \flagged{} query and $\sfrac{1}{\eta}$ total queries as above.

\paragraph{A further optimization: early stopping}
Greedy attacks such as \rays repeatedly check whether a new search direction $\theta' \gets \theta + \delta$ improves upon the current adversarial distance $\dist$, and only if so issue a call to \getdist to compute the new distance $\dist' < \dist$.
For these attacks to progress, it may not be necessary to compute $\dist'$ \emph{exactly}. Instead, knowing that $\dist' \ll \dist$ may be sufficient to know that the new direction $\theta'$ is ``\nonflagged'' and the attack can proceed with it.
We could thus stop a line search early when $\dist' \leq \gamma\cdot \dist$---for some $\gamma < 1$.
In many cases, this lets us call $\getdist$ while incurring \emph{no \flagged{} query at all}, at the expense of a less accurate distance computation.

\subsection{Stealthy Variants of Decision-based Attacks}
\label{ssec:our_attacks}

We now design stealthy variants of prior decision-based attacks, by applying the toolkit of stealthy search procedures outlined above, and illustrated in \Cref{fig:linesearch}.

\paragraph{Stealthy distance computations}
The most obvious way to make existing attacks more stealthy is to instantiate every call to $\getdist$ with a (k-stage) line search instead of a binary search. In contrast, calls to $\checkadv$ on arbitrary directions $\theta'$ are hard to make more stealthy.
This change applies to the boundary distance computation in \rays, to the gradient-estimation queries in \opt, and to the step-size searches and boundary projections in \hsj, \signopt, and \opt.
Since \ba only calls \checkadv, it cannot easily be made more stealthy.

\paragraph{Stealthy gradients}
Attacks like \opt, \signopt and \hsj use most of their queries for estimating gradients. The main difference is that instead of calling \checkadv, \opt uses more expensive calls to \getdist to get a better estimation. Prior work shows that this tradeoff is suboptimal in terms of total queries.
However, the extra precision comes for free when we consider the cost in \flagged{} queries! Recall that \checkadv yields two bits of information per \flagged{} query, while \getdist with a line-search yields $\log \sfrac{1}{\eta}$ bits.
Thus, \opt's gradient estimator is strictly better if we consider \flagged{} queries.
In \Cref{subsec:theory} we formally prove that (under mild conditions) \opt's gradient estimator gives \emph{quadratically better convergence rates} (in terms of \flagged{} queries) than the gradient estimators of \signopt and \hsj.
We can leverage this insight to design stealthy ``hybrid'' attacks that combine \opt's stealthy gradient estimator with efficient components of other, newer attacks. Specifically, we design a \stealthyhsj attack, which directly plugs in \opt's gradient estimator into \hsj's otherwise more efficient attack design. (applying this change to \signopt would just yield back the \opt attack since both attacks only differ in the gradient estimator).

\paragraph{Stealthy hyper-parameters}
Prior attacks were designed with the goal of minimizing the \emph{total} number of queries. As a result, their hyper-parameters were also tuned for this metric.
When considering our asymmetric query cost, existing hyper-parameters might thus no longer be optimal.

\paragraph{Our attacks}
We combine the above principles to design stealthy variants of existing attacks:

\begin{itemize}
	\item \textbf{\stealthyrays}: As in the original attack, in each iteration, we first greedily check if a new search direction improves the boundary distance and then replace the binary search for the new distance by a (k-stage) line-search, optionally with early-stopping (see \Cref{ssec:eggdrop}).

	\item \textbf{\stealthyopt}: The \opt attack is perfectly amenable to stealth as it only calls \getdist. We replace the original binary search with a (k-stage) line search in each of these distance computations.

	      When computing distances in random directions for estimating gradients, we need to select a safe starting point for the line search. If the current boundary distance is $\dist$, we start the search at the point at distance $(1+\gamma)\cdot \dist$ along $\theta'$, for $\gamma > 0$. If this point is not misclassified (i.e., the query is \flagged), we return $(1+2\cdot\gamma)\cdot \dist$ as an approximate distance. If the point is misclassified (i.e., safe), we perform a line search with tolerance $\sfrac{1}{\eta}$.

	      We use $\gamma=1\%$ in all experiments.

	\item \textbf{\stealthyhsj}: In each iteration, we use line searches to compute the current boundary distance and the updated step-size. We replace the original coarse-grained
	      gradient estimator (which calls \checkadv $n$ times) with \stealthyopt's estimator (with $\sfrac{n}{20}$ \getdist calls).

	\item \textbf{\stealthysignopt}: We make the same changes as for \stealthyhsj, except that we retain the original coarse-grained gradient estimator (otherwise this would be the same as \stealthyopt).
	      To better balance the number of \flagged{} queries used in different attack phases,
	      we reduce the number $n$ of queries used to estimate gradients.
	      This change is sub-optimal if we care about the attack's \emph{total} number of queries, but is beneficial in terms of \emph{\flagged} queries as the attack now spends a larger fraction of work on queries that \emph{can} be made more stealthy.
\end{itemize}

\subsection{Convergence Rates of Stealthy Attacks}\label{subsec:theory}
Before we evaluate our stealthy attacks empirically, we show that our techniques lead to attacks that are \emph{provably} more stealthy than prior attacks (under similar assumptions as considered in prior work).

Prior work has analyzed the convergence rate of SGD with the zero-order gradient estimation schemes used in \signopt and \opt \citep{liu2018zeroth,cheng2019sign}. We can use these results to prove that the gradient estimation of our \stealthyopt attack is asymptotically more efficient (in terms of \flagged{} queries) than the non-stealthy gradient estimation used by \signopt and \hsj.

Let $g(\theta)$ be the distance to the boundary along the direction $\theta$, starting from some example $x$ (this is the function that \opt and \signopt explicitly minimize).
Suppose we optimize $g$ with black-box gradient descent, using the following two gradient estimators:

\begin{itemize}
	\item \opt:  $\frac{1}{Q} \sum_{i=1}^{Q} \left(g(\theta + r_i) - g(\theta)\right) \cdot r_i$ for $Q$ random Gaussian directions $r_i$.

	\item \signopt:  $\frac{1}{Q'} \sum_{i=1}^{Q'} \sign(g(\theta + r_i) - g(\theta)) \cdot r_i$ for $Q'$ random  Gaussian directions $r_i$.
\end{itemize}

We can then show the following results:

\begin{theorem}[Adapted from \citet{liu2018zeroth} (Theorem~2)]
	\label{thm:opt}
	Assume $g$ has gradients that are $L$-Lipschitz and bounded by $C$ (assume $L$ and $C$ are constants for simplicity).
	Let $d$ be the data dimensionality.
	Optimizing $g$ with $T$ iterations of gradient descent, using \opt's gradient estimator, yields a convergence rate of $\E[\|\nabla g(x)\|_2^2] = \mathcal{O}({d}/{T})$, with $\mathcal{O}({T^2}/{d})$ \flagged{} queries.
\end{theorem}

\begin{theorem}[Adapted from \citet{cheng2019sign} (Theorem 3.1)]
	\label{thm:signopt}
	Assume $g$ is $L$-Lipschitz and has gradients bounded by $C$ (assume $L$ and $C$ are constants for simplicity).
	Let $d$ be the data dimensionality.
	Optimizing $g$ with $T$ iterations of gradient descent, using \signopt's gradient estimator, yields a convergence rate of $\E[\|\nabla g(x)\|_2] = \mathcal{O}(\sqrt{d / T})$, with $\mathcal{O}(T^2d)$ \flagged{} queries.
\end{theorem}

The convergence rate of \opt is thus at least as good as that of \signopt,%
\footnote{Note that \citet{cheng2019sign} provide a bound on the gradient norm, while \citet{liu2018zeroth} provide a bound on the \emph{squared} gradient norm.
	Applying Jensen's inequality to the result of \Cref{thm:signopt}, we know that for \signopt we have $\E[\|\nabla g(x)\|_2^2] \geq (\E[\|\nabla g(x)\|_2])^2 = \mathcal{O}(d / T)$.}
but \opt's gradient estimator with line searches requires a factor $d^2$ fewer \flagged{} queries.
The same asymptotic result as for \signopt holds for the similar estimator used by \hsj.

\begin{proof}[Proof of Theorem 1]
	\citet{liu2018zeroth} show that \opt's gradient estimator yields a convergence rate of $\E[\|\nabla g(x)\|_2^2] = \mathcal{O}(\nicefrac{d}{T}) + \mathcal{O}(\nicefrac{1}{Q})$ (see Theorem 2 in \citet{liu2018zeroth}). To balance the two convergence terms, we set $Q = \nicefrac{T}{d}$. To perform $Q$ evaluations of $g(\theta + r_i) - g(\theta)$, we need $Q+1$ calls to \getdist. Each call makes multiple queries to the model $f$, but only one \flagged{} query if we use a line search.
	This yields the number of \flagged{} queries in the theorem ($T$ iterations with $\frac{T}{d}$ \flagged{} queries per iteration).
\end{proof}

\begin{proof}[Proof of Theorem 2]
	\citet{cheng2019sign} show that \signopt's gradient estimator yields a convergence rate of $\E[\|\nabla g(x)\|_2] = \mathcal{O}(\sqrt{\nicefrac{d}{T}}) + \mathcal{O}\left(\nicefrac{d}{\sqrt{Q'}}\right)$ (see Theorem 3.1 in \citet{cheng2019sign}). To balance the two convergence terms, we set $Q' = Td$. To perform $Q'$ evaluations of $\sign(g(\theta + r_i) - g(\theta))$, one call to \getdist and $Q'$ calls to \checkadv are required. Each \checkadv call makes a single query to the model $f$, i.e., $\nicefrac{1}{2}$ \flagged{} queries on average.
	This yields the number of \flagged{} queries in the theorem ($T$ iterations with $\frac{Td}{2}$ \flagged{} queries per iteration).
\end{proof}

\section{Evaluation}
\label{sec:eval}

\begin{figure*}[t]
	\centering
	\begin{subfigure}[t]{0.29\textwidth}
		\centering
		\includegraphics[height=\plotheight]{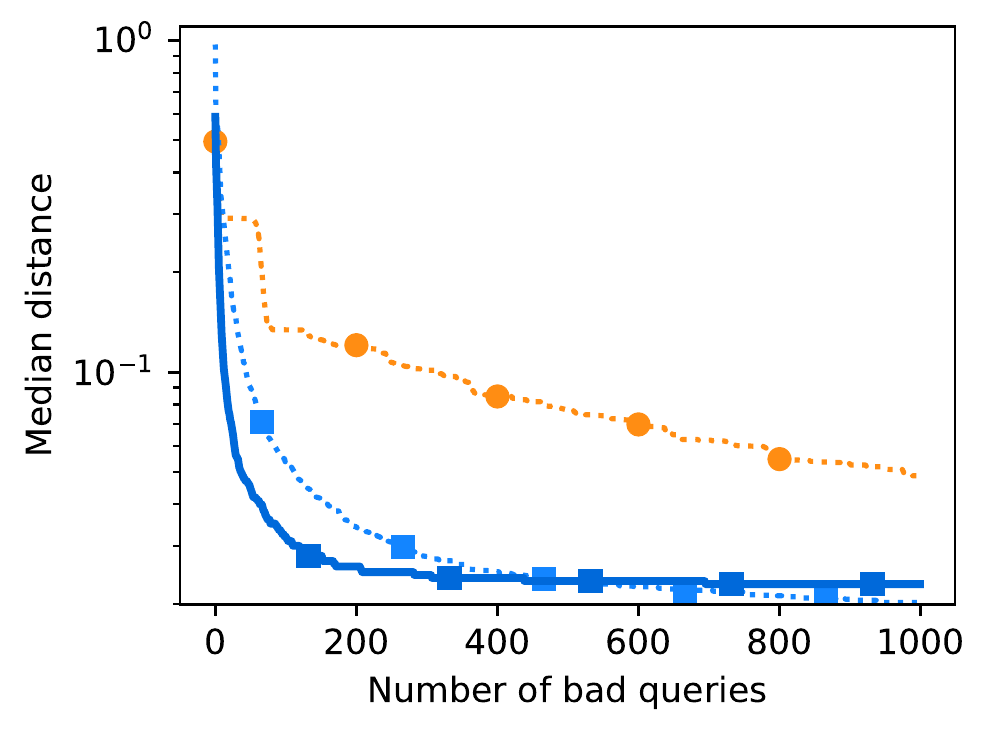}
		\caption{\footnotesize{\emph{ImageNet ($\ell_\infty$)}}}
		\label{fig:imagenet_linf}
	\end{subfigure}
	\begin{subfigure}[t]{0.29\textwidth}
		\centering
		\includegraphics[height=\plotheight]{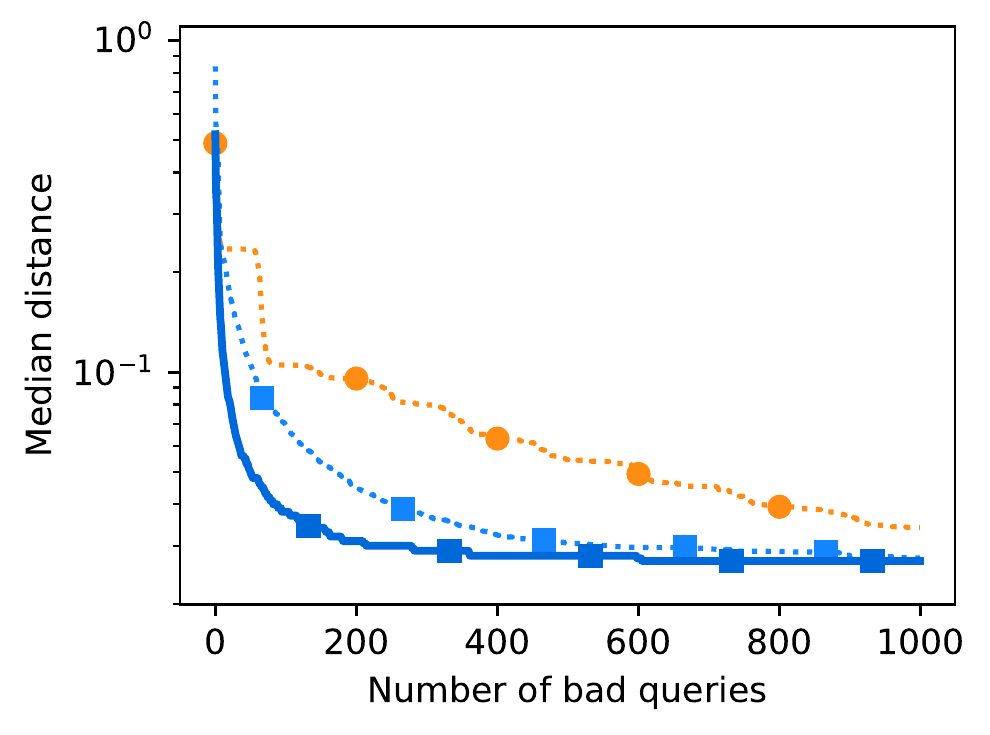}
		\caption{\footnotesize{\emph{ImageNet-Dogs ($\ell_\infty$)}}}
		\label{fig:binary_linf}
	\end{subfigure}
	\begin{subfigure}[t]{0.405\textwidth}
		\centering
		\includegraphics[height=\plotheight]{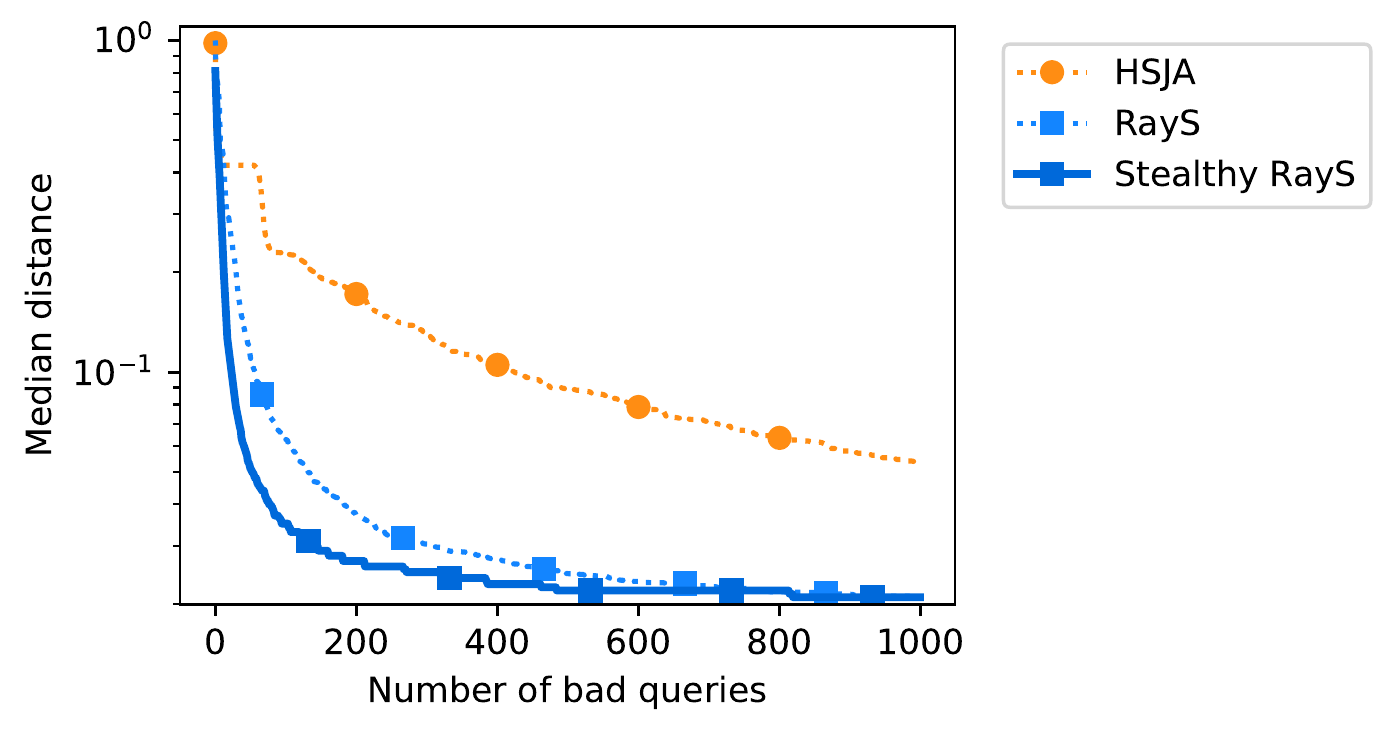}
		\caption{\footnotesize{\emph{ImageNet-NSFW ($\ell_\infty$)}}}
		\label{fig:nsfw_linf}
	\end{subfigure}
	\\[.25em]
	\begin{subfigure}[t]{0.29\textwidth}
		\centering
		\includegraphics[height=\plotheight]{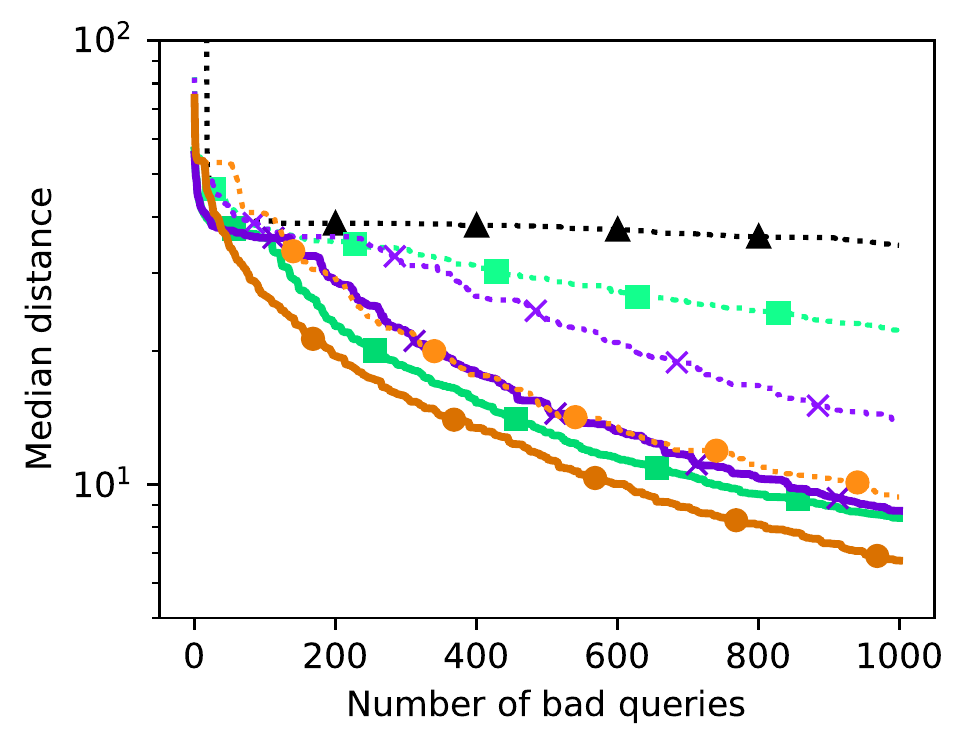}
		\caption{\footnotesize{\emph{ImageNet ($\ell_2$)}}}
		\label{fig:imagenet_l2}
	\end{subfigure}
	\begin{subfigure}[t]{0.29\textwidth}
		\centering
		\includegraphics[height=\plotheight]{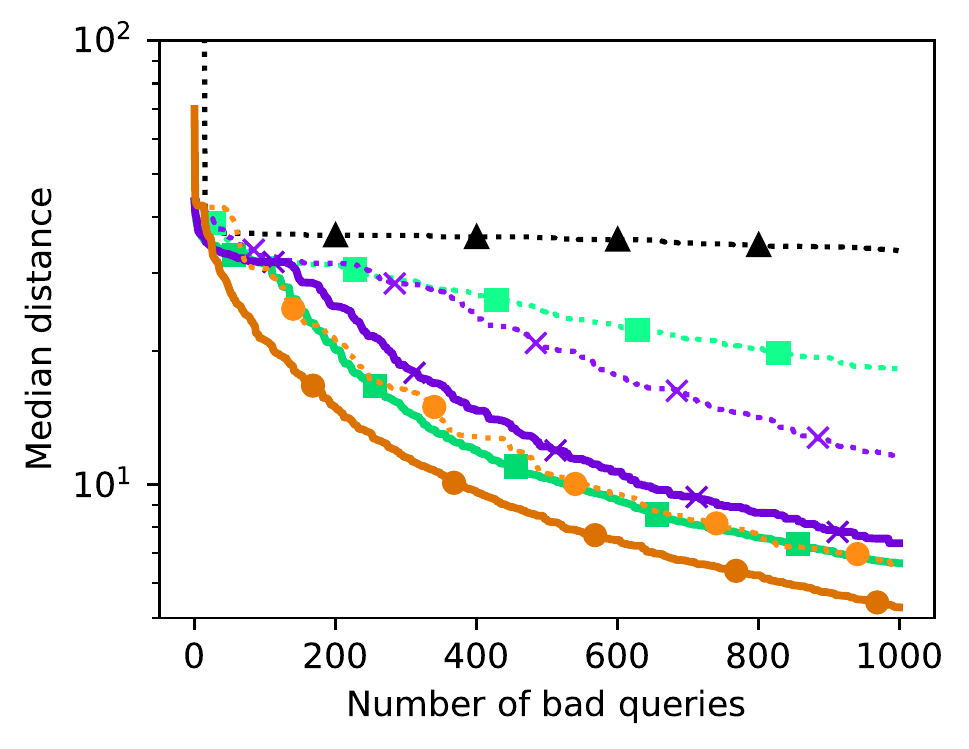}
		\caption{\footnotesize{\emph{ImageNet-Dogs ($\ell_2$)}}}
		\label{fig:binary_l2}
	\end{subfigure}
	\begin{subfigure}[t]{0.405\textwidth}
		\centering
		\includegraphics[height=\plotheight]{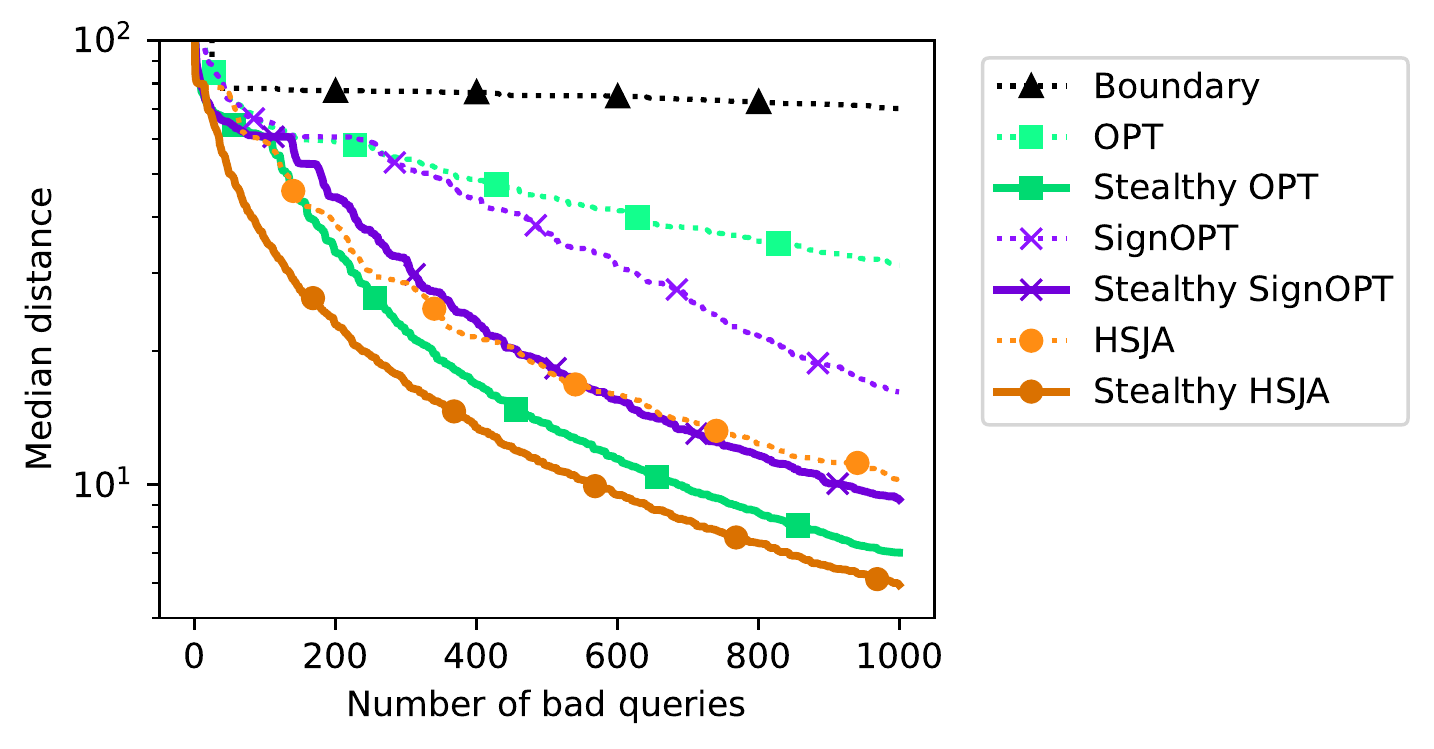}
		\caption{\footnotesize{\emph{ImageNet-NSFW ($\ell_2$)}}}
		\label{fig:nsfw_l2}
	\end{subfigure}
	\caption{Our stealthy attacks find small adversarial perturbations with fewer \flagged{} queries. For each benchmark, we report the median adversarial distance as a function of \flagged{} queries for various $\ell_2$ attacks (top) and $\ell_\infty$ attacks (bottom). Our attack variants (full lines)--designed to be stealthy--require fewer \flagged{} queries than the original attacks (dashed lines) to reach the same adversarial distance.}
	\label{fig:results}
\end{figure*}

We evaluate our stealthy decision-based attacks on a variety of benchmarks, in order to show that our attacks can drastically reduce the number of \flagged{} model queries compared to the original attacks.

\subsection{Setup}
\label{ssec:eval_setup}

\paragraph{Datasets and models}
We consider four benchmarks:

\begin{itemize}
	\item We begin with standard untargeted attacks on ImageNet against a ResNet-50 classifier.
	      We mark a query as \flagged{} if it is classified into the class of the original input.

	\item To capture more realistic security-critical scenarios, we consider a variety of binary classification tasks that aim to separate ``\nonflagged'' from ``\flagged'' data. As a toy benchmark, we use a binary labeling of ImageNet (hereafter ImageNet-Dogs), with all dog breeds grouped as the ``\flagged'' class. The classifier is also a ResNet-50, with a binary head finetuned over the ImageNet training set.

	\item We then consider an NSFW classification task with a CLIP classifier that was used to sanitize the LAION dataset~\citep{schuhmann2022laion}. To avoid collecting a new NSFW dataset, we use a subset of ImageNet (hereafter ``ImageNet-NSFW'') that this classifier labels as NSFW with high confidence.\footnote{We do not collect a new NSFW dataset due to the ethical hazards that arise from curating such sensitive data. By using a subset of ImageNet---the most popular image dataset in machine learning research---we mitigate, but do not completely eliminate \citep{prabhu2020large}, the potential harms of constructing an NSFW dataset.}

	\item Finally, we evaluate a black-box commercial NSFW detector, using our ImageNet-NSFW dataset. The detector returns a score from 1 to 5, denoting that the input is ``highly unlikely'' to ``highly likely'' to contain adult content or nudity. We consider a query to be \flagged{} if it gets a score of 4 or 5.
\end{itemize}

We provide more details on datasets and models in Appendix~\ref{apx:setup}.

\paragraph{Attacks} We evaluate \hsj and \rays for $\ell_\infty$ attacks, and \ba, \opt, \signopt and \hsj for $\ell_2$ attacks.
We adapt each attack's official code to enable the counting of \flagged{} queries. We use each attack's default hyper-parameters, except for some optimizations by \citet{sitawarin2022preprocessors} (see Appendix~\ref{apx:setup}).

We further evaluate our \textit{stealthy} versions (i.e., designed to be stealthy) of \opt, \signopt, \hsj, and \rays.
For \stealthyopt, \stealthysignopt, and \stealthyhsj we
split search intervals into $10{,}000$ sub-intervals, and perform either a full line search or a two-stage line search with $100$ coarse-grained and fine-grained steps. For efficiency's sake, we perform two-stage line searches in all our experiments and use the results to infer the number of queries incurred by a full line search.
For \stealthysignopt, we further trade-off the query budgets for computing gradients and step-sizes by reducing the attack's default number of gradient queries $n$ by a factor $k\in \{1.5, 2.0, 2.5\}$, and find the $k$ (which we call ``optimal $k$'') that provides the best results. For the step-size searches, we use the same line-search procedure as in \opt.

For \stealthyrays, we replace each binary search with a line-search of step-size $\eta=10^{-3}$ (the default binary search tolerance for \rays) and implement early-stopping with $\gamma=0.9$.

\paragraph{Metrics} As in prior work, we report the median $\ell_p$ norm of adversarial examples after $N$ attack queries (except we only count \emph{\flagged} queries).
For each task, we run the attacks on $500$ samples from the corresponding test set (for ImageNet-Dogs, we only attack images of dogs).
For the attacks on the commercial NSFW detector, use use $200$ samples from ImageNet-NSFW.

Our motivation for counting \flagged{} queries is to assess whether black-box attacks are viable for attacking real security systems. We thus focus on a ``low'' query regime: each attack can make at most $1{,}000$ \flagged{} queries per sample. Prior work has considered much larger query budgets, which we disregard here as such budgets are likely not viable against systems that implement any query monitoring.

\subsection{Results}\label{ssec:eval_results}

The main results of our evaluation appear in \Cref{fig:results}. We also provide a full ablation over different attack variants and optimizations in \Cref{tab:ablation}.
For all benchmarks, our stealthy attacks (with 1-stage line searches) issue significantly fewer \flagged{} queries than the corresponding original attack.

\begin{figure*}[t]
	\centering
	\begin{subfigure}[t]{0.29\textwidth}
		\centering
		\includegraphics[height=\plotheight]{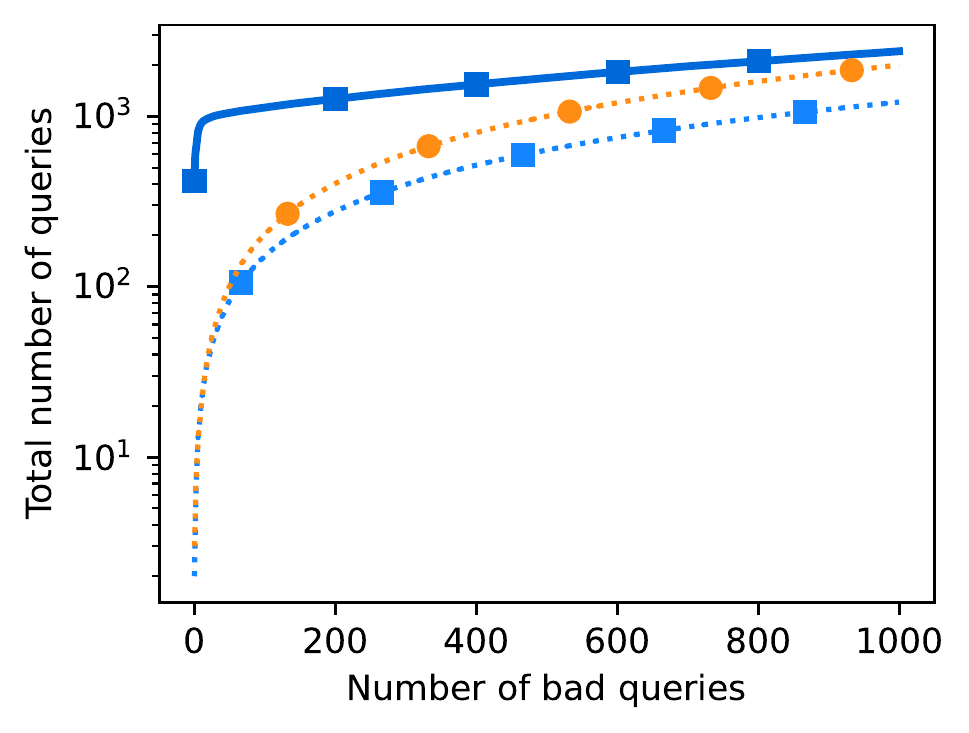}
		\caption{\footnotesize{\emph{ImageNet ($\ell_\infty$)}}}
	\end{subfigure}
	\begin{subfigure}[t]{0.29\textwidth}
		\centering
		\includegraphics[height=\plotheight]{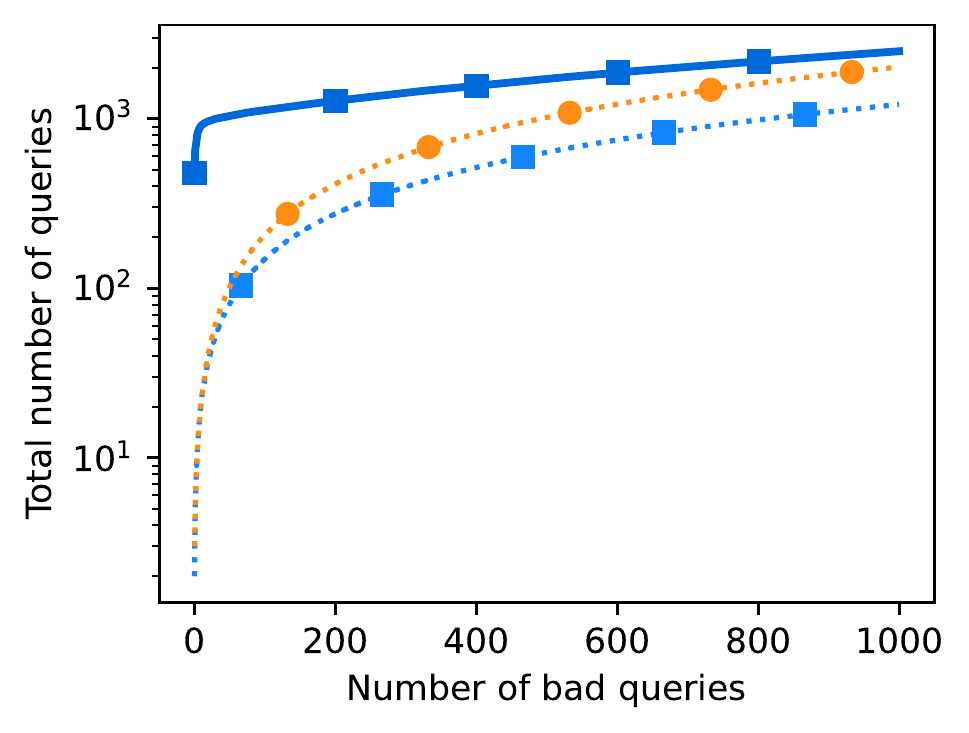}
		\caption{\footnotesize{\emph{ImageNet-Dogs ($\ell_\infty$)}}}
	\end{subfigure}
	\begin{subfigure}[t]{0.405\textwidth}
		\centering
		\includegraphics[height=\plotheight]{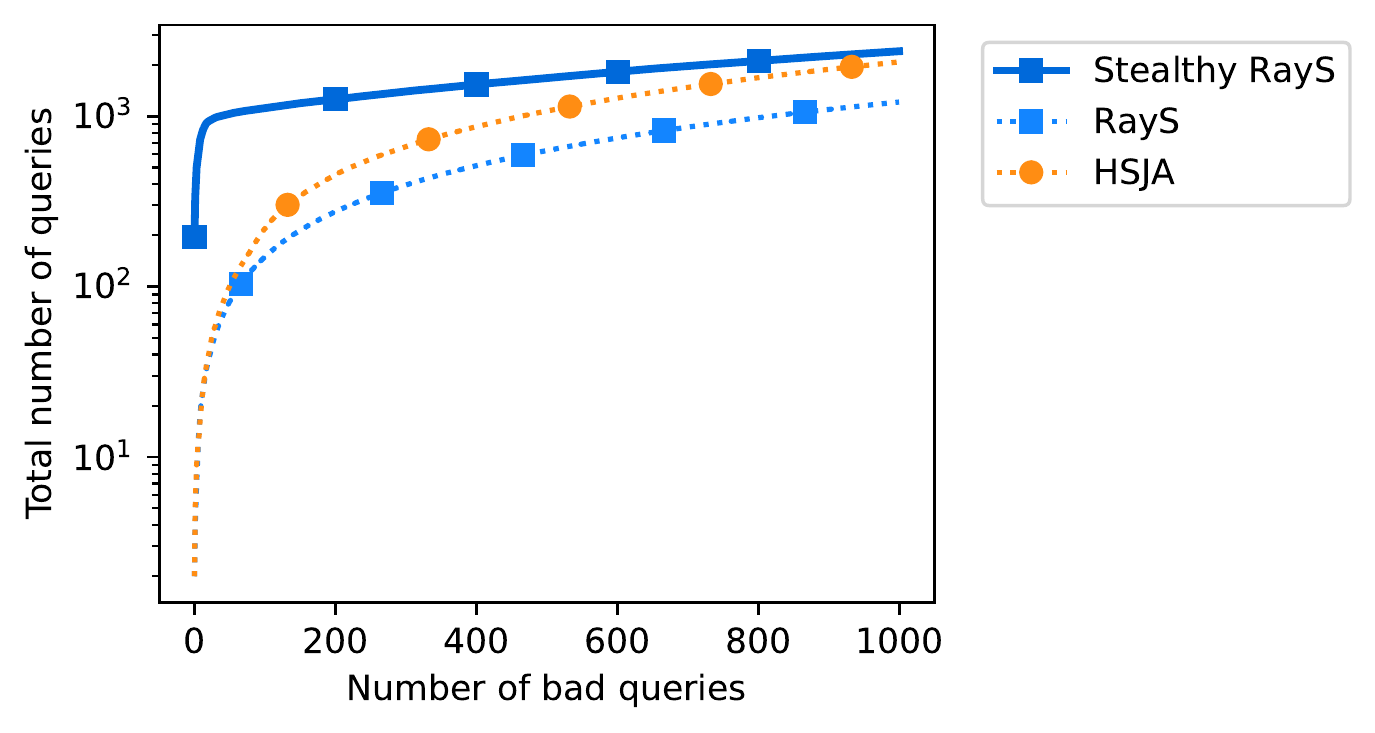}
		\caption{\footnotesize{\emph{ImageNet-NSFW ($\ell_\infty$)}}}
	\end{subfigure}
	\\[.25em]
	\begin{subfigure}[t]{0.29\textwidth}
		\centering
		\includegraphics[height=\plotheight]{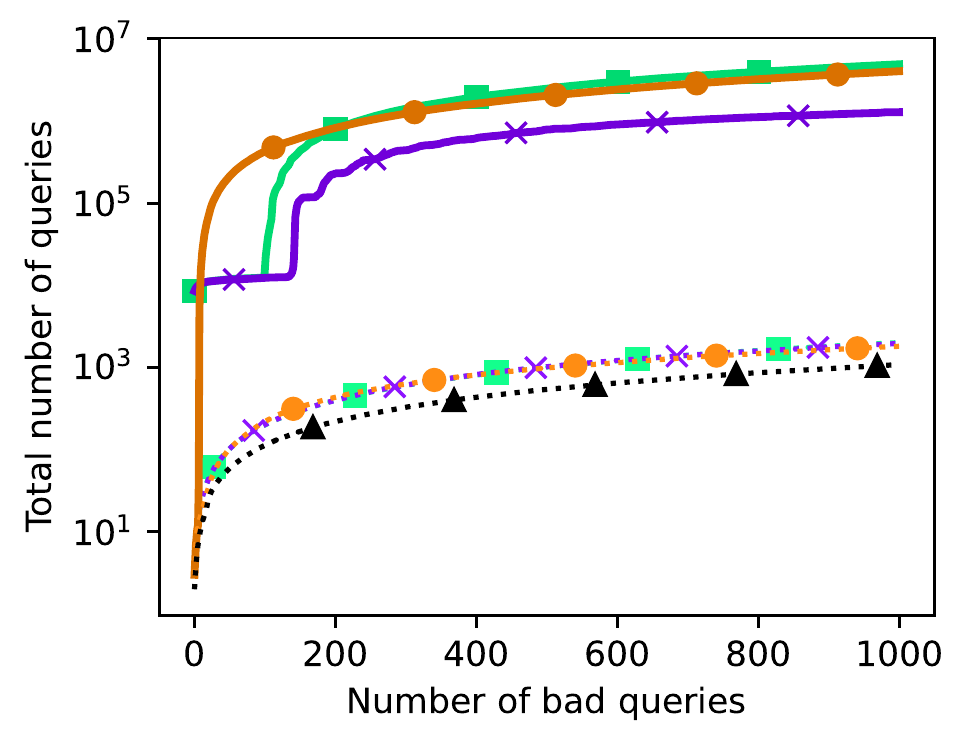}
		\caption{\footnotesize{\emph{ImageNet ($\ell_2$)}}}
	\end{subfigure}
	\begin{subfigure}[t]{0.29\textwidth}
		\centering
		\includegraphics[height=\plotheight]{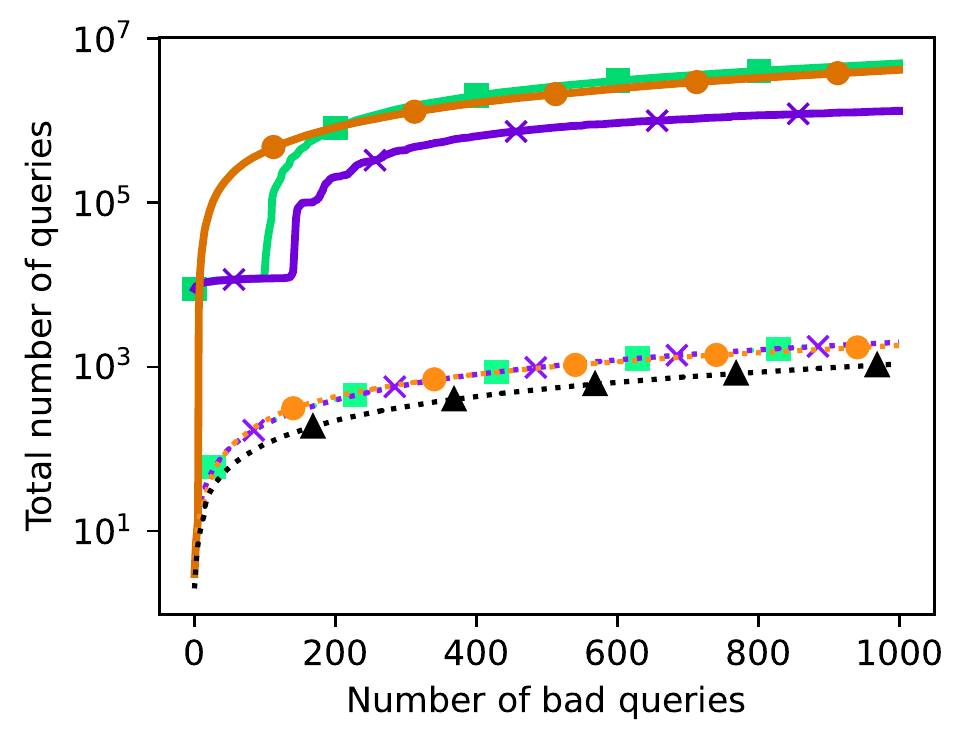}
		\caption{\footnotesize{\emph{ImageNet-Dogs ($\ell_2$)}}}
	\end{subfigure}
	\begin{subfigure}[t]{0.405\textwidth}
		\centering
		\includegraphics[height=\plotheight]{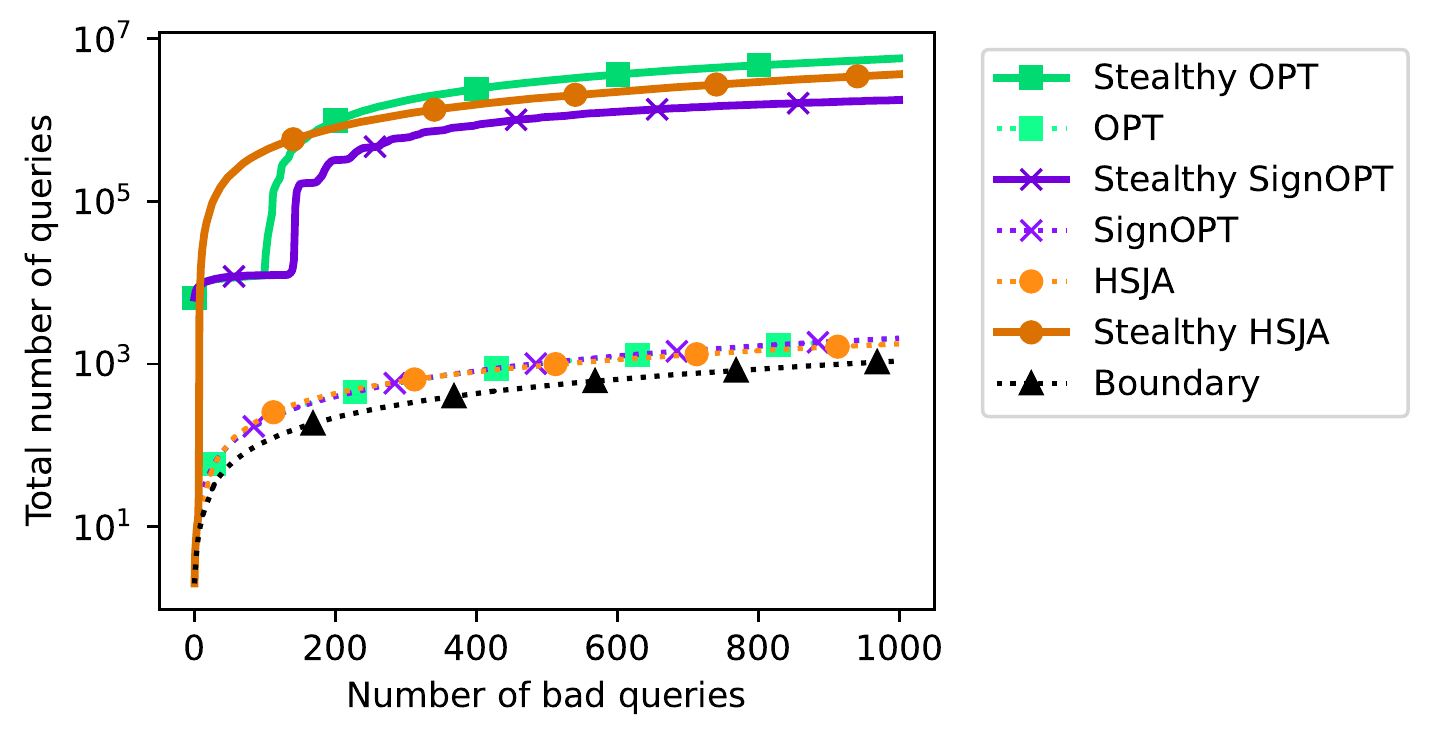}
		\caption{\footnotesize{\emph{ImageNet-NSFW ($\ell_2$)}}}
	\end{subfigure}
	\caption{Trade-offs between total queries and \flagged{} queries made by different attacks. Our stealthy attacks (full lines) issue many more queries than their original counterparts (dashed lines).}
	\label{fig:total_queries}
\end{figure*}

\paragraph{Stealthy $\ell_\infty$ attacks are cost-effective}
Our \stealthyrays attack reduces \flagged{} queries compared to the original \rays, which is itself more efficient than \hsj.
To reach a median norm of $\epsilon=\sfrac{8}{255}$, \stealthyrays needs $103$--$181$ \flagged{} queries for the three benchmarks, $2.1$--$2.4\times$ less than \rays, and $7$--$17\times$ less than \hsj.
As \stealthyrays issues only $2.1$--$3.4\times$ more queries than \rays (see \Cref{fig:total_queries}), it is clearly cost effective if $c_{\textrm{\flagged}} \gg c_0$.

Due to the greedy nature of \rays, it fails to find very small perturbations. \hsj thus outperforms \rays given a large enough query budget. But this regime is likely unimportant for practical attacks, both because the query budget required is too high, and because perturbations of the size found by \rays are likely sufficient in practice.

\begin{figure}[t]
	\centering
	\includegraphics[width=0.8\columnwidth]{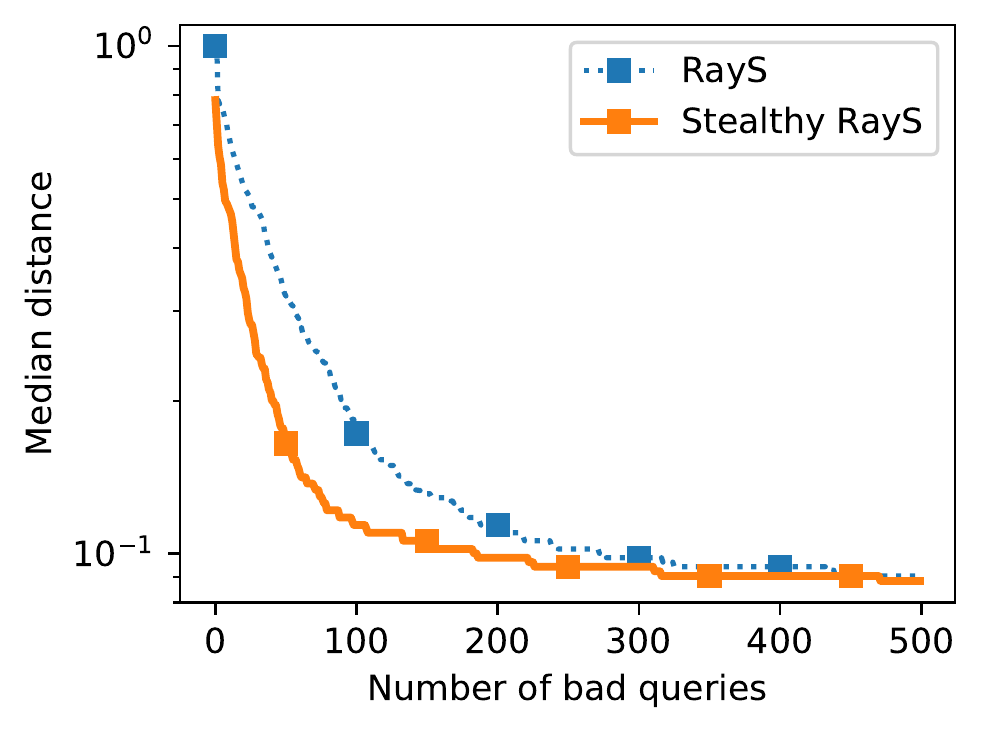}
	\caption{Attack on a commercial black-box NSFW detector. We run \rays and \stealthyrays on 200 samples from our ImageNet-NSFW dataset. We denote a query as \flagged{} if it is flagged as ``likely'' to be NSFW. The stealthy attack needs $2.2\times$ fewer \flagged{} queries to find adversarial perturbations of size $\epsilon=\sfrac{32}{355}$.}
	\label{fig:google_nsfw}
\end{figure}

\paragraph{Attacking a real black-box NSFW detector}\label{ssec:eval_nsfw}
We now turn to a much more realistic attack scenario where we target a commercial black-box detector of NSFW images using \stealthyrays. Few decision-based attacks in the literature have been evaluated against real black-box ML systems. In fact, many attacks require non-trivial changes to work against a real ML system which expects queries to be valid 8-bit RGB images.
Notably, the binary-search tolerance $\eta$ used in the attacks we have considered is orders of magnitude smaller than the minimal distance between two RGB images.
The few attacks that have been evaluated against commercial systems (e.g., the \ba, or \textsc{Qeba}~\citep{li2020qeba}) used a limited number of attack samples (3 to 5) due to the high query cost---and thus monetary cost---of evaluating these attacks against a commercial API.
To enable a more rigorous evaluation, we focus here on \rays---the only attack we evaluated that reliably finds small adversarial perturbations on a limited query budget (\textless 500 queries).

Since real black-box systems expect 8-bit RGB images as input, we set \rays's threshold $\eta$ for a binary search or line-search to $\sfrac{1}{255}$, the smallest distance between two distinct RGB images. This is much coarser than the default threshold of $\eta=10^{-3}$, and the attack thus finds  larger perturbations.
Recall that in each iteration \rays checks whether a small change to the current direction results in a smaller $\ell_\infty$ perturbation. The issue is that with discretized images, the smallest measurable change in the $\ell_\infty$ norm is $\sfrac{1}{255}$. Thus, the attack only works if small changes to the adversarial direction result in significant reductions of the $\ell_\infty$ norm.
Other decision-based attacks face similar quantization issues when applied to real black-box systems. We thus encourage future work to take into account query discretization when designing black-box attacks.

\begin{figure}[t]
	\centering
		\begin{subfigure}[t]{\columnwidth}
            \centering
			\includegraphics[width=0.8\columnwidth]{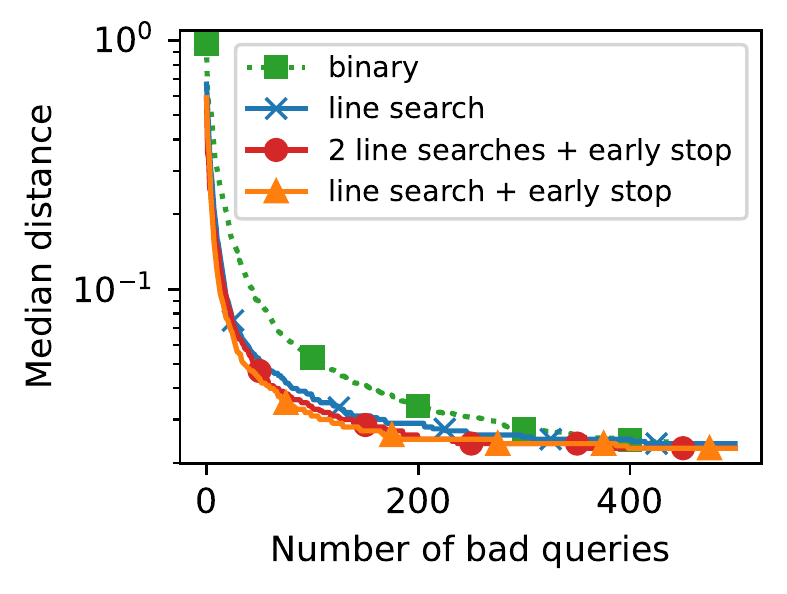 }
		\end{subfigure}
		\\
        \begin{subfigure}[t]{\columnwidth}
            \centering  
			\includegraphics[width=0.8\columnwidth]{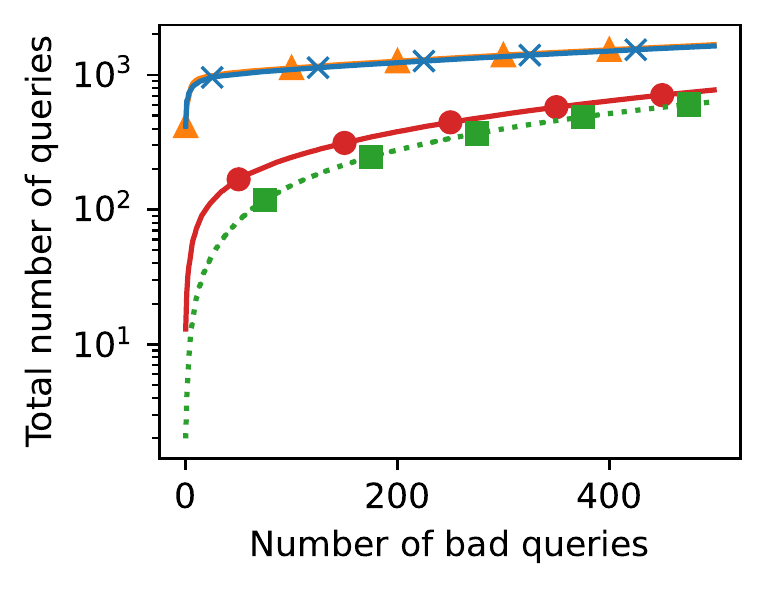 }
		\end{subfigure}
		\caption{Trade-offs between \nonflagged{} and \flagged{} queries for various search strategies in the \rays attack on ImageNet.}\label{fig:rays_tradeoff}
\end{figure}

We evaluate \rays and \stealthyrays on 200 images from ImageNet-NSFW. The API that we attack naturally expects to be queried with NSFW data and thus does not use any asymmetric pricing of queries (i.e., all API queries have the same cost). But we of course still distinguish \flagged{} queries (classified as NSFW) from \nonflagged{} ones, as such \flagged{} queries would incur a large cost when attacking a real application that makes use of an NSFW detector model.
\Cref{fig:google_nsfw} shows the results. Evading this commercial detector is much harder than the prior models we attacked, presumably due to the discretization constraint described above. %
Our \stealthyrays attack outperforms \rays by $2.2\times$ (we reach a median distance of $\sfrac{32}{255}$ with $79$ \flagged{} queries, while \rays needs $172$ \flagged{} queries). These perturbations are noticeable, but preserve the images' NSFW nature. This comes at a reasonable overhead in terms of overall queries, which is $1.21\times$ higher for \stealthyrays.

\paragraph{Stealthy $\ell_2$ attacks are more cost-effective than non-stealthy ones, but likely impractical}
Remarkably, while \opt is one of the earliest and least efficient decision-based attacks, our \stealthyopt variant is stealthier than the newer \signopt and \hsj attacks.
To reach a median $\ell_2$ perturbation of $10$ on ImageNet, \stealthyopt needs $686$ \flagged{} queries, a saving of $7.3\times$ over the original \opt, and of $1.4\times$ compared to \hsj.
Our hybrid \stealthyhsj attack is the stealthiest attack overall.
On all three benchmarks, it requires $1.47$--$1.82\times$ fewer \flagged{} queries than \hsj to reach a median perturbation of $10$.
This shows that we can even improve the stealthiness of attacks that do not make use of many distance queries. Our techniques are thus likely also applicable to other decision-based attacks that follow \hsj's blueprint.

\Cref{fig:total_queries} shows the \emph{total} number of queries made by our stealthy attacks. As expected, our stealthy attacks issue many more queries in total than attacks that optimize for this quantity. To reach a median perturbation of $\epsilon=10$, our attacks make $350$--$1420\times$ more total queries than the original non-stealthy attack.
This large increase is only warranted if benign queries are significantly cheaper than \flagged{} queries. This may be the case in some applications, e.g., uploading $1{,}000$ benign images is permitted on platforms like Facebook~\citep{facebook_albums}, and thus likely less suspicious than a \emph{single} \flagged{} query. However, for less extreme asymmetries in query costs (e.g., $c_{\text{\flagged}} = 10 \cdot c_0$), a less strict tradeoff between \flagged{} and \nonflagged{} queries is warranted. We will explore this in \Cref{ssec:tradeoff}.

In \Cref{fig:costs}, we further show the total $\texttt{cost}$ of our attacks for various configurations of the query costs $c_0$ and $c_{\textrm{\flagged}}$. A different attack variant is optimal depending on the cost overhead of \flagged{} queries.

\begin{table*}[t]
        \normalsize
	\caption{Ablation on stealthy attack components. For attacks on ImageNet, we show the relative number of \flagged{} queries and \nonflagged{} queries (lower is better) compared to the original non-stealthy attack, to achieve a median $\ell_2$ perturbation of 10, or $\ell_\infty$ perturbation of $\sfrac{8}{255}$.}
	\label{tab:ablation}
	\centering
        \setlength{\tabcolsep}{4pt}
	\begin{tabular}{@{}l lcc@{}}
		\textbf{Attack}                             & \textbf{Ablation}                                           & \specialcell[b]{\textbf{\Flagged{} queries reduction}                              \\\textbf{(higher is better)}}  & \specialcell[b]{\textbf{\Nonflagged{} queries increase}\\\textbf{(lower is better)}} \\ \toprule
		\multirow{2}{*}{{\hsj}}     &
		with OPT grad estimation       & $1.56\times$                                       & $1418.72\times$                                                           \\
		                                   & with OPT grad estimation + 2-stage line search & $0.78\times$                                 & $\hphantom{00}30.05\times$ \\ \midrule
		\multirow{2}{*}{{\opt}}     &
		with line search                   & $7.25\times$                                       & $\hphantom{0}351.96\times$                                                \\
		                                   & with 2-stage line search                           & $4.09\times$                                 & $\hphantom{000}4.62\times$ \\ \midrule
		\multirow{4}{*}{{\signopt}} &
		with optimal $k$                     & $1.06\times$                                       & $\hphantom{000}0.93\times$                                                \\
		                                   & with line search                                   & $1.26\times$                                 & $\hphantom{0}393.60\times$ \\
		                                   & with line search + optimal $k$                       & $1.81\times$                                 & $\hphantom{0}386.60\times$ \\
		                                   & with 2-stage line search + optimal $k$               & $1.77\times$                                 & $\hphantom{000}4.79\times$ \\ \midrule
		\multirow{3}{*}{{\rays}}    &
		with line search                   & $1.98\times$                                       & $\hphantom{000}3.42\times$                                                \\
		                                   & with line search + early stopping                  & $2.37\times$                                 & $\hphantom{000}3.42\times$ \\
		                                   & with 2-stage line search + early stopping          & $1.77\times$                                 & $\hphantom{000}0.92\times$ \\ \bottomrule
	\end{tabular}
\end{table*}

\subsection{Trading off \Nonflagged and \Flagged{} Queries}
\label{ssec:tradeoff}

\begin{figure*}[h]
	\centering
	\begin{subfigure}[b]{0.33\textwidth}
		\includegraphics[height=3.5cm]{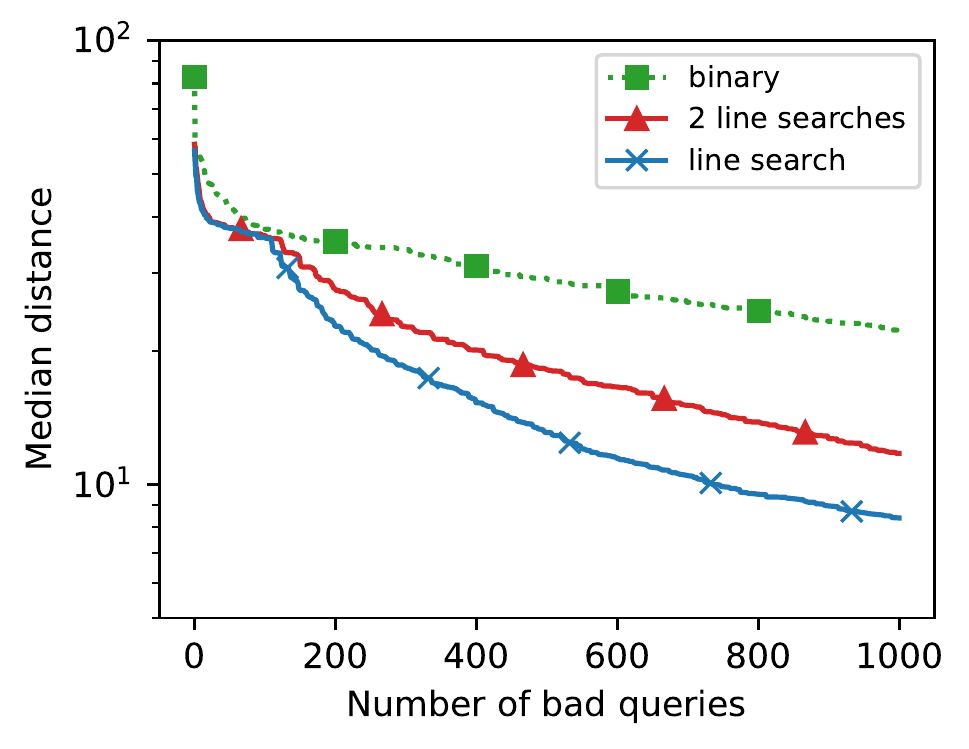 }
	\end{subfigure}
	\quad
	\begin{subfigure}[b]{0.33\textwidth}
		\includegraphics[height=3.5cm]{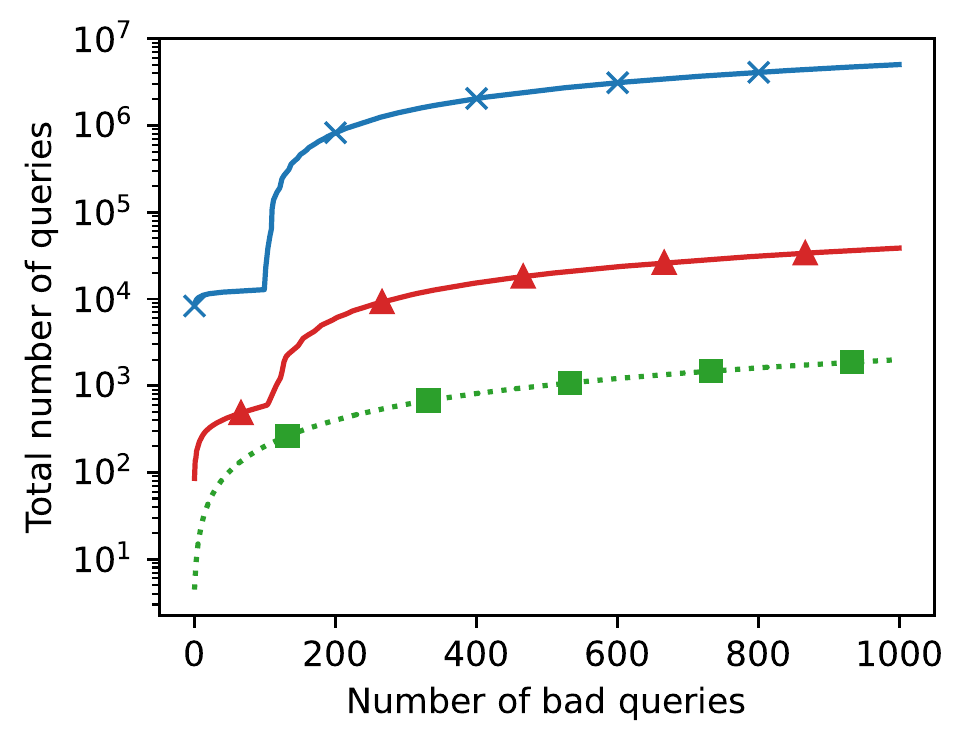 }
	\end{subfigure}
	\caption{Trade-offs between \nonflagged{} and \flagged{} queries for different search strategies in the \stealthyopt attack. A full line search makes one \flagged{} query and up to $10{,}000$ \nonflagged{} queries. The version with two searches makes two \flagged{} queries and up to $2\cdot 100$ \nonflagged{} queries.}
	\label{fig:opt_tradeoffs}
    \vspace{-1em}
\end{figure*}

Our stealthy attacks in \Cref{fig:results} use full line searches, which use a \emph{single} \flagged{} query (and many \nonflagged{} queries). In \Cref{fig:rays_tradeoff} and \Cref{fig:opt_tradeoffs} we consider alternative tradeoffs. We provide a full ablation over different attack variants and optimizations in \Cref{tab:ablation}.

For $\ell_\infty$ attacks, \stealthyrays with a two-stage line-search and early stopping provides a nice tradeoff: for a median perturbation of $\epsilon=\sfrac{8}{255}$, the attack makes $1.37\times$ more \flagged{} queries than a full line-search, but $3.7\times$ fewer total queries. This attack is actually \emph{strictly better} than the original \rays (thanks to early stopping): our attack makes $1.77\times$ fewer \flagged{} queries, and $8\%$ fewer \nonflagged{} queries!

For $\ell_2$ attacks, \stealthyopt with a two-stage line-search %
shows a nice tradeoff over the original \opt: for a median perturbation of $\epsilon=10$, our attack makes $4\times$ fewer \flagged{} queries, at the expense of $5\times$ more \nonflagged{} queries (see \Cref{fig:opt_tradeoffs}).
Unfortunately, none of our stealthy attacks with two-stage line searches beat the original \hsj in terms of \flagged{} queries. Thus, attaining state-of-the-art stealthiness with our techniques does appear to come at the expense of a large overhead in \nonflagged{} queries.
As a result, improving the total cost of existing $\ell_2$ decision-based attacks may be hard, and thus attacking real security-critical systems with these attacks may simply not be cost-effective.

\section{How asymmetric are query costs in practice?}\label{sec:costs}

Whether the trade-offs provided by stealthy attacks are worthwhile is application-dependent. Security critical platforms such as social media websites or App stores provide few details about their filtering systems, for obvious reasons.
Nevertheless, some platforms such as Facebook or Twitter ($\mathbb{X}$) do publish moderation policies that allow us to offer an educated guess on the relative costs of flagged and non-flagged queries in real-world systems ($c_{\textrm{\flagged}}$ and $c_0$, respectively). We use Facebook and Twitter as case-studies below.

According to Meta, an account will ``get a 1-day restriction from creating content'' after seven violations of the ``Community Standards'', and a ``a 30-day restriction from creating content'' after ten violations. Stronger measures, up to account termination, are taken for further violations~\citep{facebook_bans}.
At the same time, users can upload up to $1{,}000$ photos per album~\citep{facebook_albums}, suggesting that thousands of \nonflagged pictures can be uploaded without raising suspicion.
Twitter's policies suggest similar numbers: users can create up to $2{,}400$ posts per day~\citep{twitter_limits}; accounts get suspended after ``repeated violations'' (i.e., \flagged queries)~\citep{twitter_bans}, which we speculate to mean 5--10 violations as in the case of Facebook.
Given these numbers, we posit that an adversary could make at least three to four orders of magnitude more \nonflagged queries than \flagged queries with a single account. Under the assumption that the cost of setting up new accounts is the predominant attacker cost---and that a successful attack should not take more than a few days to run---we conclude that $c_0 / c_{\textrm{\flagged}}$ is on the order of $10^{-3}$ or $10^{-4}$.

\begin{figure*}[t]

	\centering
	\begin{subfigure}[t]{0.29\textwidth}
		\centering
		\includegraphics[height=\plotheight]{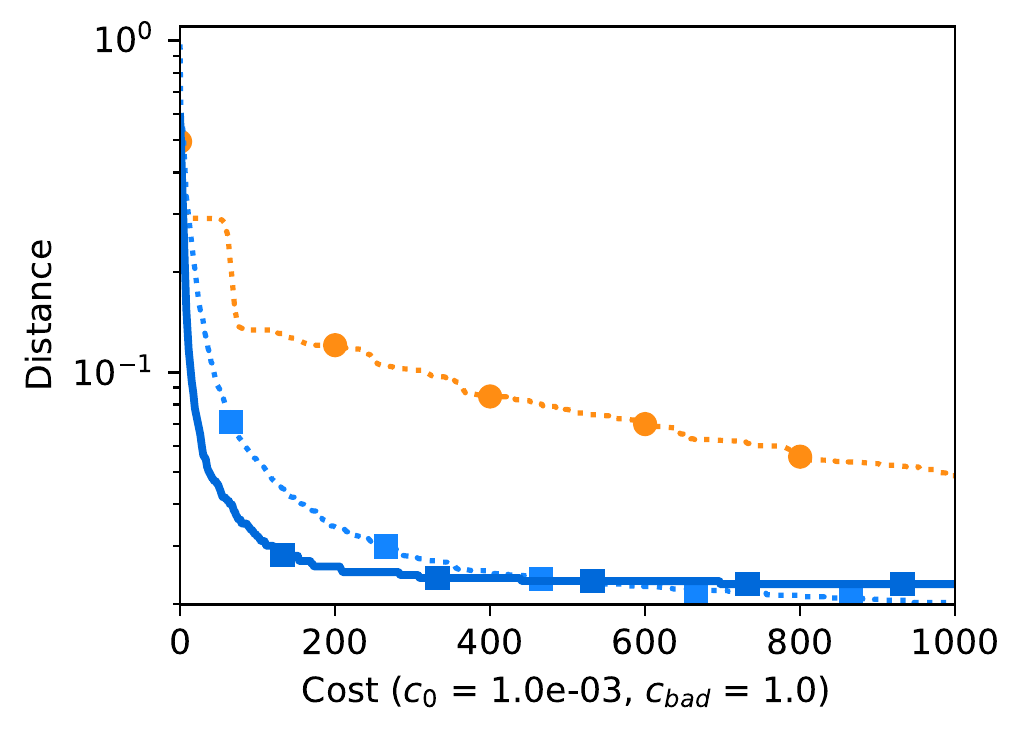}
		\caption{\footnotesize{$c_0 = 10^{-3}$ ($\ell_\infty$)}}
		\label{fig:imagenet_linf_cost_1e-3}
	\end{subfigure}
	\begin{subfigure}[t]{0.29\textwidth}
		\centering
		\includegraphics[height=\plotheight]{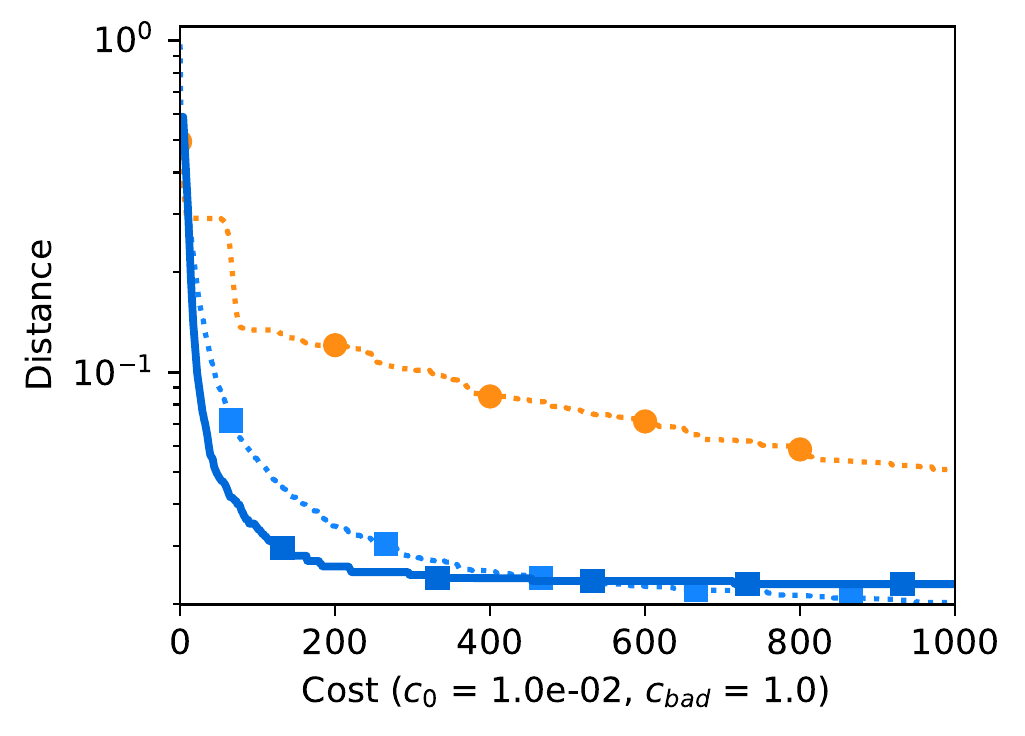}
		\caption{\footnotesize{$c_0 = 10^{-2}$ ($\ell_\infty$)}}
		\label{fig:imagenet_linf_cost_1e-2}
	\end{subfigure}
	\begin{subfigure}[t]{0.405\textwidth}
		\centering
		\includegraphics[height=\plotheight]{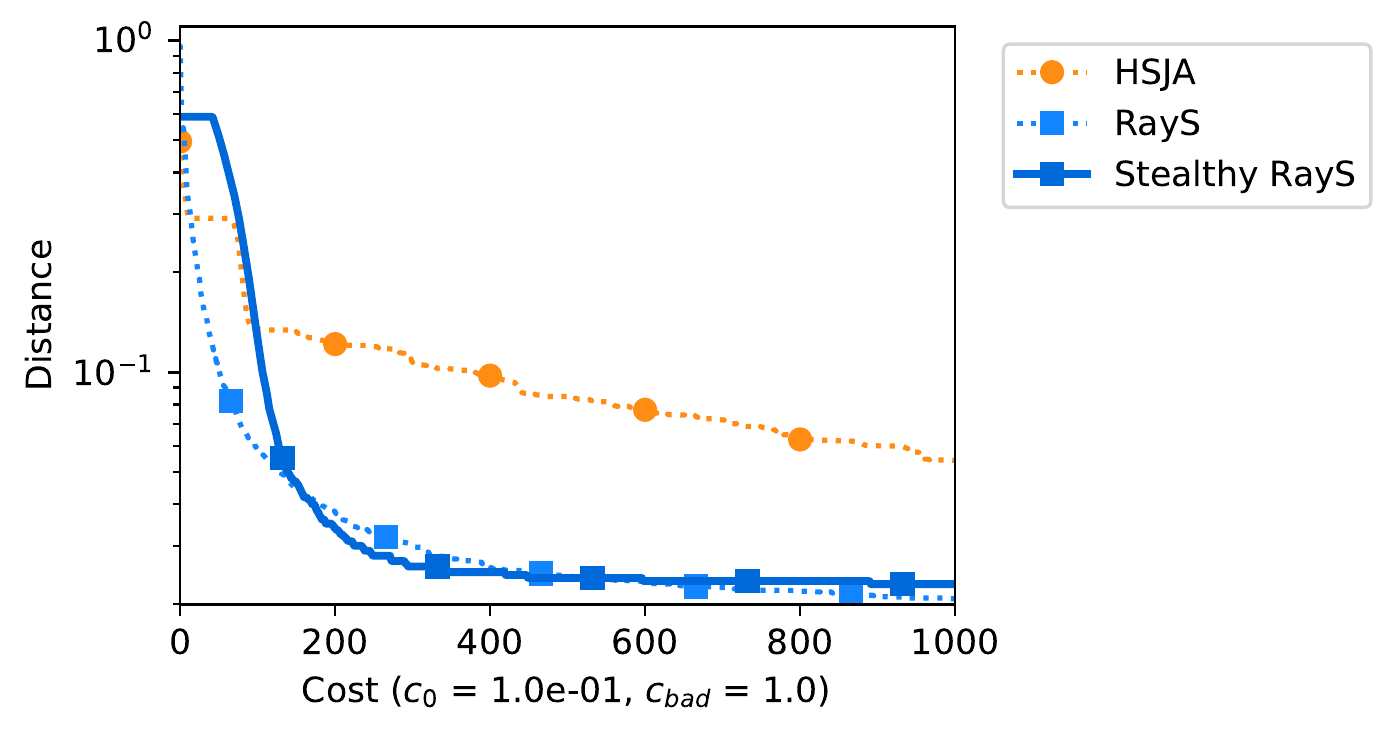}
		\caption{\footnotesize{$c_0 = 10^{-1}$ ($\ell_\infty$)}}
		\label{fig:imagenet_linf_cost_1e-1}
	\end{subfigure}
	\\[.25em]
	\begin{subfigure}[t]{0.29\textwidth}
		\centering
		\includegraphics[height=\plotheight]{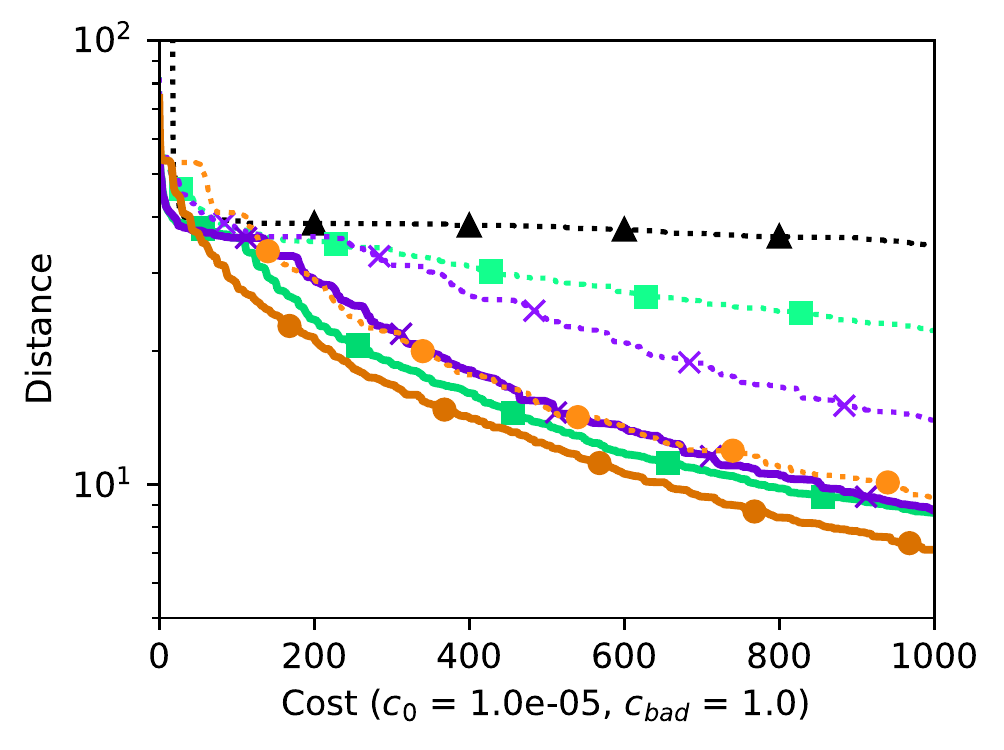}
		\caption{\footnotesize{$c_0 = 10^{-5}$ ($\ell_2$)}}
		\label{fig:imagenet_l2_cost_1e-5}
	\end{subfigure}
	\begin{subfigure}[t]{0.29\textwidth}
		\centering
		\includegraphics[height=\plotheight]{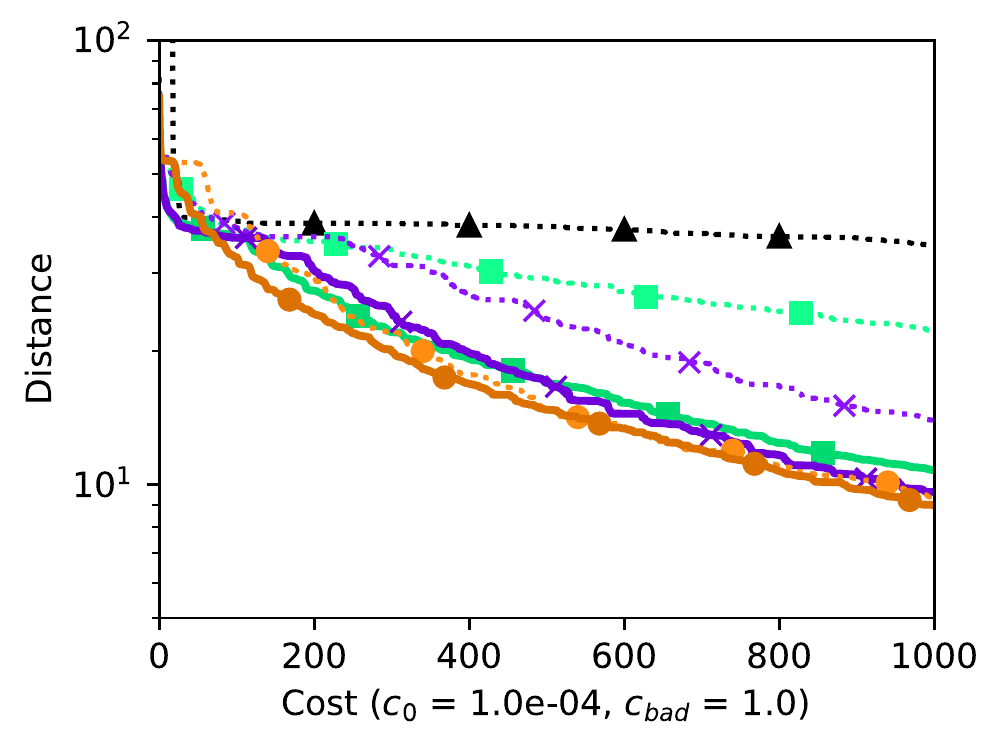}
		\caption{\footnotesize{$c_0 = 10^{-4}$ ($\ell_2$)}}
		\label{fig:imagenet_l2_cost_1e-4}
	\end{subfigure}
	\begin{subfigure}[t]{0.405\textwidth}
		\centering
		\includegraphics[height=\plotheight]{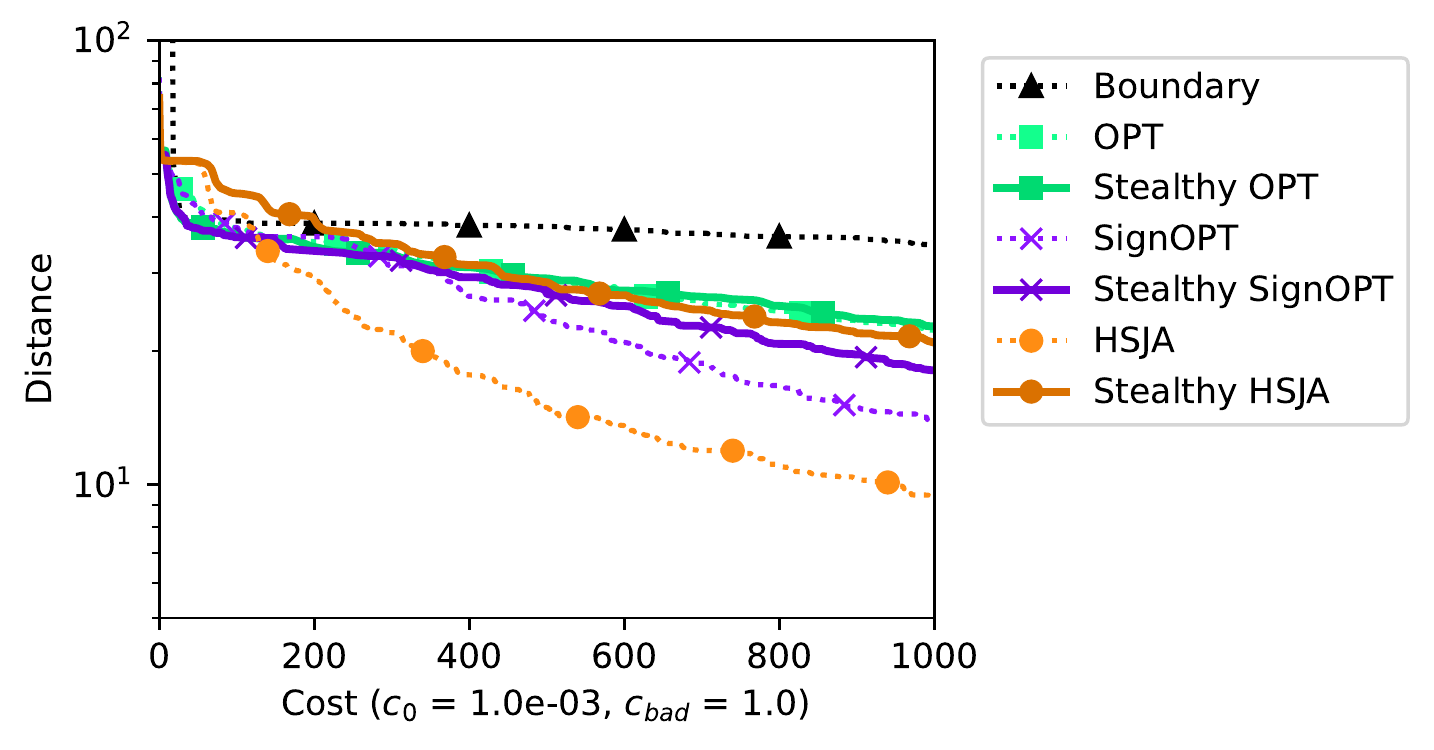}
		\caption{\footnotesize{$c_0 = 10^{-3}$ ($\ell_2$)}}
		\label{fig:imagenet_l2_cost_1e-3}
	\end{subfigure}
	\caption{Costs trade-offs of various decision-based attacks on ImageNet, for different asymmetric costs of \nonflagged{} and \flagged{} queries. We show how the attack cost varies for different values of the base query cost $c_0$, at a fixed cost $c_{\textrm{\flagged}}=1$ for bad queries. The advantage given by the stealthy attacks is reduced when the relative cost of good queries increases.}
	\label{fig:costs}
\end{figure*}

In this regime, our $\ell_\infty$ \stealthyrays attack is clearly more cost efficient than prior $\ell_\infty$ attacks (as we can see from \Cref{fig:imagenet_linf_cost_1e-3}).
For $\ell_2$ attacks the situation is less clear. It is possible that our \stealthyhsj attack would be cost effective (over prior attacks) in some settings, but further attack improvements are likely necessary to obtain a stealthy attack that could realistically be applied against a deployed system.

To illustrate the practicality of the tradeoff offered by our \stealthyrays attack, we go back to our experiment on attacking a commercial NSFW detection API from \Cref{ssec:eval_nsfw}.
Suppose the attacker wants to evade detection with a perturbation of at most $\epsilon=\sfrac{32}{355}$.
Our \stealthyrays attack needs $311$ overall queries, of which $79$ are \flagged, while \rays needs $225$ queries, of which $172$ are flagged.
For a platform that uses a moderation policy similar to Meta's, a stealthy attacker would require setting up $\sfrac{79}{7}\simeq 12$ accounts. In contrast, the non-stealthy attacker would need $\sfrac{172}{7}\simeq25$ accounts, i.e., $2.1\times$ more.
Note that in this regime, neither attacker ever hits the moderation limits for unflagged queries, and thus the attack cost is purely a function of the number of flagged queries (which our stealthy attacks explicitly minimize).

\section{Related Work}\label{sec:related_work}

\paragraph{Threat models for ML evasion attacks}
Modeling realistic ML evasion attacks is challenging~\citep{gilmer2018motivating, apruzzese2022real}. Our work contributes to this goal by introducing the more realistic \emph{asymmetric query cost} metric and evaluating the feasibility of stealthy decision-based attacks.
Prior work has attacked real security-critical ML systems such as malware detectors~\citep{biggio2013evasion}, copyright systems~\citep{saadatpanah2020adversarial}, or online content blockers~\citep{tramer2018adblock, yuan2019stealthy}. These works either assume white-box model access, or use black-box \emph{transfer} attacks. Decision-based attacks against real ML systems have been mounted against systems that expose an explicit query API (and which do not appear to monitor queries for inappropriate data)~\citep{brendel2017decision, li2020qeba, zheng2021black}.

\paragraph{Transfer-based evasion attacks}\label{par:transfer}
Transfer-based evasion attacks leverage the fact that an input that is an adversarial example for a ``surrogate'' white-box model could also be a successful adversarial example for another, black-box model~\citep{papernot2016transferability}. They are perfectly stealthy (they make no \flagged{} queries to the target black-box model) but have limited success rates, as they are not model-specific, unlike decision-based attacks. The fact that they have limited success rate does not affect the performance of the attacks under our metric of counting \flagged queries: transfer-based attacks make no queries at all. Moreover, previous work also explored the possibility of combining transfer with decision-based attacks by using transfer-based priors~\citep{brunner2019guessing}. Our metric is still useful for the decision-based component of these attacks.
 
 A further possibility to leverage transferability could be to train (or fine-tune) a surrogate model based on the outputs of the black-box model targeted by the adversary. However, to achieve this, the adversary needs to label enough training samples (both \flagged and \nonflagged) with the target model, to train the surrogate model. To achieve this, they would need to query the model with several \flagged samples. This presents the same cost-asymmetry issue as existing decision-based attacks. While we believe that studying the amount of \flagged queries needed to train such a model is out of the scope of this work, our metric would still be useful when performing this kind of evaluation.

\paragraph{Detecting decision-based attacks}
\citet{chen2020stateful} and \citet{li2022blacklight} detect decision-based attacks by monitoring sequences of user queries.
Our notion of stealthiness does not consider such defenses (but it could be expanded in this way if such defenses were adopted). 
We aim to evade a more fundamental form of monitoring that any security-critical system likely uses: flagging and banning users who issue many ``\flagged'' queries.

\paragraph{Stealthy score-based attacks}
\emph{Score-based attacks}, which query a model's confidence scores~\citep{chen2017zoo}, also issue many \flagged{} queries. Designing stealthy score-based attacks is similar to the problem of ``safe black-box optimization'' in reinforcement learning~\citep{usmanova2020safe}.
It is unclear whether any security-critical ML system would return confidence scores. In such a case, existing score-based attacks would be blatantly \emph{non-stealthy}, as they typically perform zeroth-order gradient ascent starting from the original input, and thus issue \emph{only} unsafe queries. 

\section{Conclusion}
Our paper initiates the study of \emph{stealthy} decision-based attacks, which minimize costly \emph{\flagged} queries that are flagged by an ML system.
Our ``first-order'' exploration of the design space for stealthy attacks shows how to equip existing attacks with stealthy search procedures, at a cost of a larger number of benign queries.
Decision-based attacks may be made even stealthier by designing them \emph{from scratch} with stealth as a primary criterion. We leave this as an open problem we hope future work can address.

We hope our paper will pave the way towards more refined analyses of the cost of evasion attacks against real ML systems.
In particular, our paper suggests a new possible defense metric for defenses designed to resist black-box attacks: the number of \flagged{} queries before an attack is effective.

\bibliography{main}
\bibliographystyle{plainnat}

\clearpage

\appendices

\section{Additional experimental details}\label{apx:setup}

\subsection{Datasets and Models}

\paragraph{ImageNet} We run the attacks against a ResNet-50~\citep{he2016deep} classifier trained on ImageNet~\citep{deng2009imagenet}. We use the model weights provided as part of the torchvision library~\citep{paszke2019pytorch}, which reach $76.13\%$ validation accuracy. When running the attacks, we use ImageNet's validation set and we skip the samples that are already classified incorrectly by the model.

\paragraph{ImageNet-Dogs} We create a binary classification task from ImageNet by considering as ``\flagged'' the images belonging to classes of dog breeds (i.e., the classes with indices included in the range $[151, 268]$) and as ``\nonflagged'' the images belonging to all the other classes. We create training and validation sets in this way from the respective splits of ImageNet.
Then, we take the ResNet-50 provided by torchvision, change the last linear layer to a layer with one output, and fine-tune this model for one epoch on the training set, using Adam~\citep{kingma2014adam} with learning rate $10^{-3}$. Training the model takes around 1 hour using an Nvidia RTX A6000. The final model has $96.96\%$ accuracy, $87.14\%$ precision, and $87.10\%$ recall on the validation set. Since we are interested in creating adversarial examples for the ``\flagged'' images, we only attack the images in the validation set that are correctly classified as ``\flagged'' (i.e., as dogs) by the fine-tuned model.

\paragraph{ImageNet-NSFW} As mentioned in \cref{ssec:eval_results}, we also evaluate the attacks on the NSFW content detector shared by \citet{schuhmann2022laion}. This classifier takes as input CLIP~\citep{radford2021learning} embeddings of images and outputs a confidence in $[0, 1]$. We use the CLIP implementation provided by the HuggingFace Transformers library~\citep{wolf2020transformers} to extract the CLIP embeddings from the input images. To create an evaluation set of NSFW images, we select the subset of 1,000 images in the ImageNet validation set that the NSFW content detector classifies as NSFW with highest confidence (it is well known that ImageNet contains NSFW content~\citep{prabhu2020large}). When attacking the model, we consider an attack to be successful if the confidence of the detector drops below $0.5$.

\subsection{Attack Hyper-parameters}

\paragraph{\ba}
We use the official implementation\footnote{\small{\url{https://github.com/bethgelab/foolbox/blob/1c55ee/foolbox/attacks/boundary_attack.py}}}, which is part of from Foolbox~\citep{rauber2017foolbox}, with default hyper-parameters on all tasks.

\paragraph{\hsj}
We use the official implementation.\footnote{\small{\url{https://github.com/Jianbo-Lab/HSJA/blob/daecd5/hsja.py}}} Following~\citet{sitawarin2022preprocessors}, we set $\texttt{gamma} = 10{,}000$ (this hyper-parameter is used to determine the binary search threshold), as this gives better results.

\paragraph{\rays and \stealthyrays}
We use the official implementation.\footnote{\small{\url{https://github.com/uclaml/RayS/blob/29bc17/RayS.py}}}
The attack has no hyper-parameters. The default binary search tolerance is $\eta=10^{-3}$.
For the line-search in \stealthyrays we use the same step size of $10^{-3}$ and perform either a full line-search or a two-stage search by first dividing the $N$ search intervals into coarse groups of size $\sqrt{N}$.
For attacking the commercial black-box NSFW classifier in \Cref{ssec:eval_nsfw}, we set the binary search tolerance and line-search step-size to $\eta=\sfrac{1}{255}$ and perform a full line-search.
For the early-stopping optimization, we end a line search if $\dist' < 0.9 \cdot \dist$.

In \Cref{fig:results}, \Cref{fig:google_nsfw} and \Cref{fig:total_queries}, the \stealthyrays attack is the version with a full line-search and early-stopping.

\paragraph{\opt and \stealthyopt}
We use the official implementation.\footnote{\small{\url{https://github.com/cmhcbb/attackbox/blob/65a82f/attack/OPT_attack.py}}} Following~\citet{sitawarin2022preprocessors}, we set $\beta = 10^{-2}$ (this hyper-parameter is used to determine the binary search threshold).%

For \stealthyopt, we do line searches for gradient estimation in the interval $[0.99\cdot \dist, 1.01\cdot \dist]$, where $\dist$ is the current adversarial distance. For computing step sizes, we do a line search in the interval $[0.99\cdot \dist, \dist]$, since we only care about the new distance if it improves upon the current one.
We split this interval into $N=10{,}000$ sub-intervals and perform a 2-stage line-search with $100$ coarse-grained steps and 100 fine-grained steps. For efficiency sake, we \emph{batch} the line-search by calling the model on two batches of size 100, one for all coarse-grained steps, and one for all fine-grained steps. To count the number of \flagged{} queries and total queries, we assume that the line-search queries were performed one-by-one. If the first query in a line search is not safe (i.e., the boundary distance is larger than $1.01\cdot \dist$, we approximate the distance by $\dist' \approx 2\cdot \dist$.

In \Cref{fig:results} and \Cref{fig:total_queries}, the \stealthyopt attack is the version with a full line search.

\paragraph{\signopt and \stealthysignopt}
We use the official implementation.\footnote{\small{\url{https://github.com/cmhcbb/attackbox/blob/65a82f/attack/Sign_OPT.py}}} Following~\citet{sitawarin2022preprocessors}, we set $\beta = 10^{-2}$ (this hyper-parameter is used to determine the binary search threshold).

For \stealthysignopt, we do the same line search procedure as \stealthyopt for computing step sizes.
We change the default number of gradient estimation queries per iteration from $n=200$ to $n/k$ for $k \in \{1.5, 2, 2.5, 3\}$, i.e., $n \in \{67, 80, 100, 133\}$.

In \Cref{fig:results} and \Cref{fig:total_queries}, the \stealthysignopt attack uses a full line-search, and $k=2.5$.

\subsection{Compute and code}

We run every attack on one Nvidia RTX 3090, and the time to run the attacks on 500 samples ranges from twelve hours, for the attacks ran with binary search, to more than three days for the slowest attacks (e.g. OPT) ran with line search. We wrap all the attack implementations in a common set-up for which we use PyTorch~\citep{paszke2019pytorch}. The code can be found at the following URL: \url{https://anonymous.4open.science/r/realistic-adv-examples-CD4C/}. The checkpoints of the model we trained, the NSFW classifier we ported from Keras to PyTorch, and the outputs of this model on the ImageNet train and validation datasets can be found at the following URL: \url{https://osf.io/bhfcj/files/osfstorage?view_only=b55a3077521242a287ba957bd461fe59}.

\clearpage

\onecolumn
\section{Additional Figures}\label{apx:figures}

\begin{figure*}[h]
	\centering
	\begin{subfigure}[b]{0.33\textwidth}
		\includegraphics[height=3.5cm]{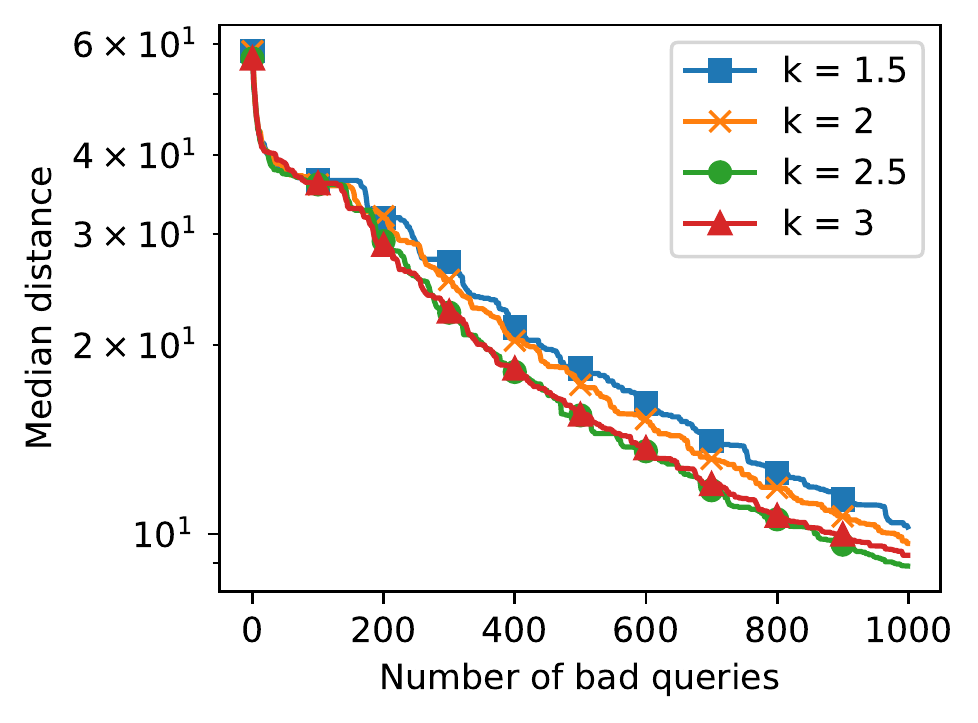 }
	\end{subfigure}
	\quad
	\begin{subfigure}[b]{0.33\textwidth}
		\includegraphics[height=3.5cm]{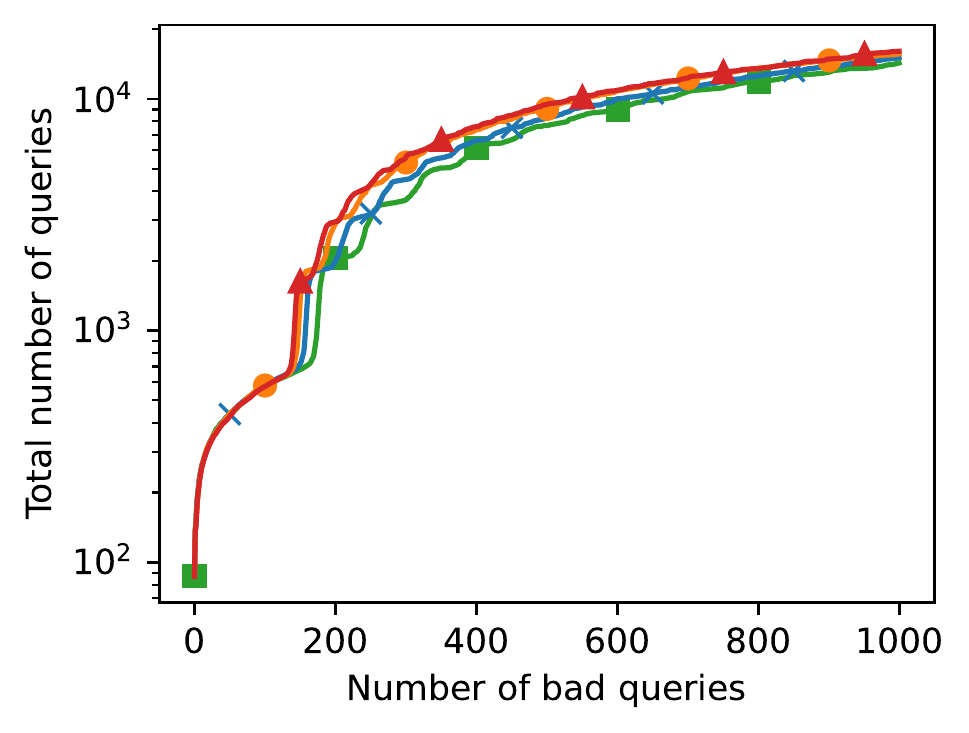 }
	\end{subfigure}
	\caption{Influence of the hyper-parameter $k$ in the \stealthysignopt attack (the reduction in the number of gradient estimation queries per iteration). The best results are obtained with $k=2.5$.}
	\label{fig:signopt_ablation}
  \vspace{-1em}
\end{figure*}

\begin{figure*}[h]
	\centering
	\begin{subfigure}[b]{0.33\textwidth}
		\includegraphics[height=3.5cm]{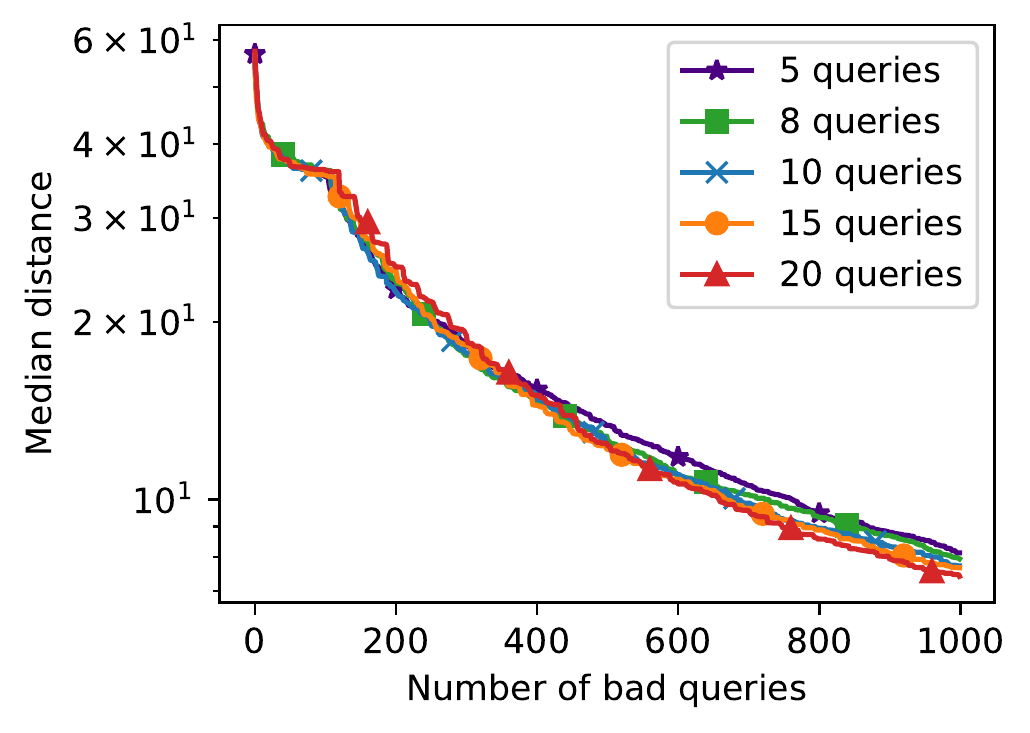 }
	\end{subfigure}
	\quad
	\begin{subfigure}[b]{0.33\textwidth}
		\includegraphics[height=3.5cm]{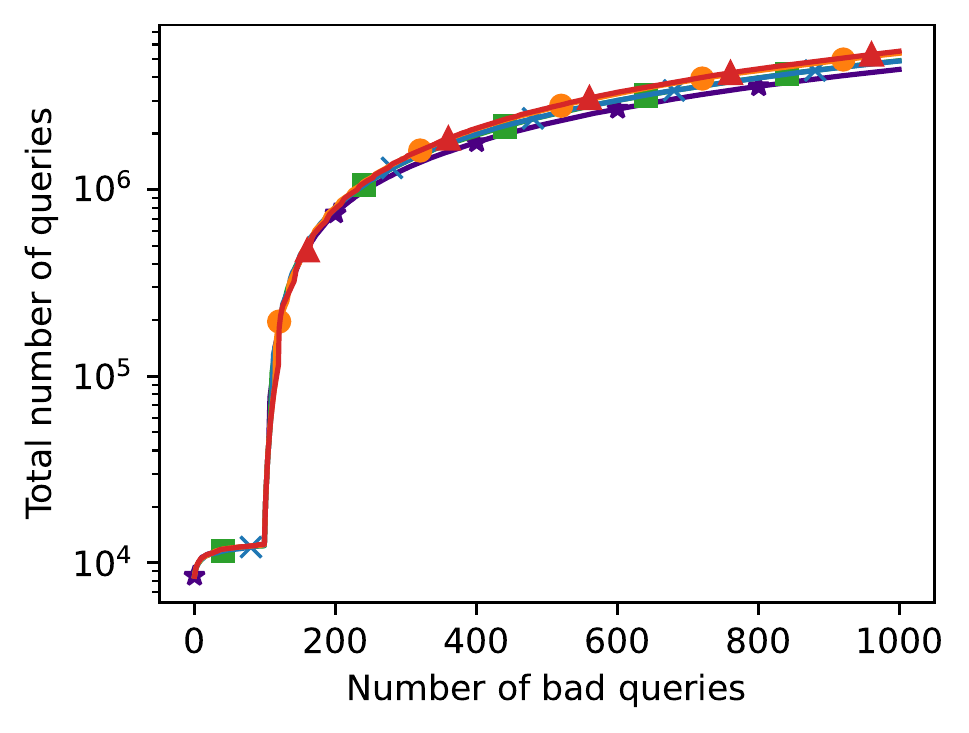 }
	\end{subfigure}
	\caption{Influence of the number of directions computed for the gradient estimation in the \stealthyopt attack. The best results are obtained with $q = 10$, which is the original value from \citet{cheng2018query}.}
	\label{fig:opt_grad_queries_ablation}
  \vspace{-1em}
\end{figure*}

\begin{figure*}[h]
	\centering
	\begin{subfigure}[b]{0.33\textwidth}
		\includegraphics[height=3.5cm]{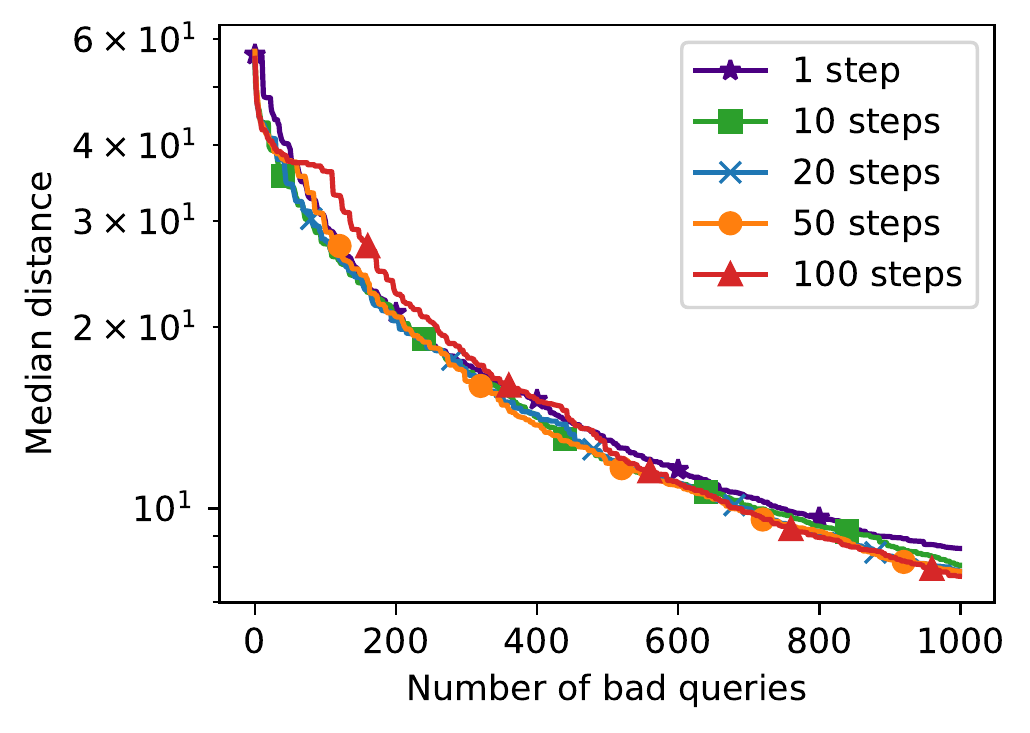 }
	\end{subfigure}
	\quad
	\begin{subfigure}[b]{0.33\textwidth}
		\includegraphics[height=3.5cm]{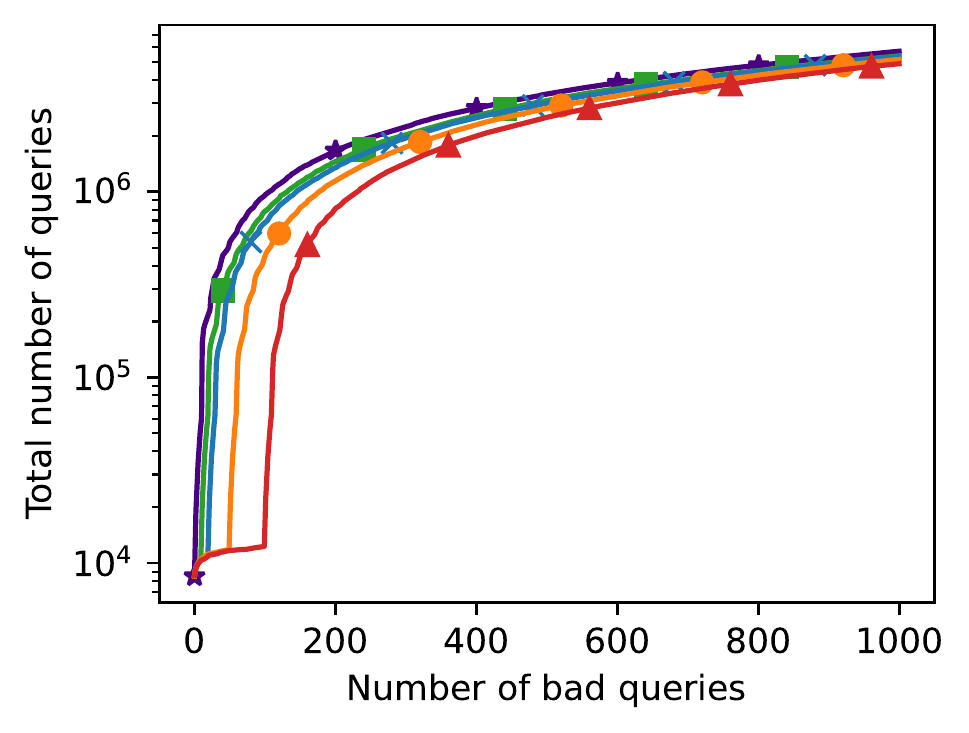 }
	\end{subfigure}
	\caption{Influence of the number of directions tested for the initialization in the \stealthyopt attack. The best results are obtained with $n=100$, when considering a larger number of queries, even though the difference between the different values is small.}
	\label{fig:opt_init_ablation}
\end{figure*}

\end{document}